\documentclass{elsart}

\usepackage{amsmath}
\usepackage{amsfonts}
\usepackage{thophys}
\usepackage{graphicx}
\usepackage{url}

\newcounter{bla}
\newenvironment{refnummer}{%
\list{[\arabic{bla}]}%
{\usecounter{bla}%
\setlength{\itemindent}{0pt}%
\setlength{\topsep}{0pt}%
\setlength{\itemsep}{0pt}%
\setlength{\labelsep}{2pt}%
\setlength{\listparindent}{0pt}%
\settowidth{\labelwidth}{[9]}%
\setlength{\leftmargin}{\labelwidth}%
\addtolength{\leftmargin}{\labelsep}%
\setlength{\rightmargin}{0pt}}}
{\endlist}

\DeclareMathOperator{\tr}{tr}
\allowdisplaybreaks[1]

\begin{document}
\begin{frontmatter}

\begin{flushright}
\begin{scriptsize}
EDINBURGH-2010-27
\\
FR-PHENO-2010-036 
\end{scriptsize}    
\end{flushright}

\title{\texttt{CleGo}: A package for automated computation of Clebsch-Gordan
 Coefficients in Tensor Products Representations \\ for Lie Algebras
 $A$-$G$} 

\author[a,b]{Christoph Horst} and
\author[a,c]{J\"urgen Reuter}

\address[a]{University of Freiburg, Institute of Physics,
  Hermann-Herder-Str. 3, 79104 Freiburg, Germany} 
\address[b]{University of Hamburg, II. Institute for Theoretical
 Physics, Luruper Chaussee 149, 22761 Hamburg, Germany}
\address[c]{University of Edinburgh, School of Physics and Astronomy,
 JCMB, The King's Buildings, Mayfield Road, Edinburgh EH9 3JZ, U.K.}

\begin{abstract}
We present a program that allows for the computation of tensor
products of irreducible representations of Lie algebras $A-G$ based on
the explicit construction of weight states. This straightforward
approach (which is slower and more memory-consumptive than the
standard methods to just calculate dimensions of the tensor product
decomposition) produces Clebsch-Gordan coefficients that are of
interest for instance in discussing symmetry breaking in model
building for grand unified theories. For that purpose, multiple tensor
products have been implemented as well as means for analyzing the
resulting effective operators in particle physics.

\end{abstract}

\end{frontmatter}

% Computer program descriptions should contain the following
% PROGRAM SUMMARY.

{\bf PROGRAM SUMMARY}
 %Delete as appropriate.

\begin{small}
\noindent
{\em Authors:} Christoph Horst, J\"urgen Reuter  \\
{\em Program Title:} \texttt{CleGo}  \\
{\em Journal Reference:}                                      \\
 %Leave blank, supplied by Elsevier.
{\em Catalogue identifier:}                                   \\
 %Leave blank, supplied by Elsevier.
{\em Licensing provisions:} none  \\
 %enter "none" if CPC non-profit use license is sufficient.
{\em Programming language:} \texttt{O'Caml}  \\
{\em Computer:} i386-i686, x86\_64  \\
 %Computer(s) for which program has been designed.
{\em Operating system:} cross-platform, for definiteness though we
assume some UNIX system.  \\
 %Operating system(s) for which program has been designed.
{\em RAM:} $\geq 4$ GB commendable, though in general memory requirements
depend on the size of the Lie algebras and the representations involved. \\
 %RAM in bytes required to execute program with typical data.
 % Please give some freely chosen keywords that we can use in a
 % cumulative keyword index.
{\em Classification:} 4.2, 11.1                                         \\
 %Classify using CPC Program Library Subject Index, see (
 % http://cpc.cs.qub.ac.uk/subjectIndex/SUBJECT_index.html)
 %e.g. 4.4 Feynman diagrams, 5 Computer Algebra.
{\em External routines/libraries:} \texttt{nums.cma} (exact integer arithmetic,
provided by the \texttt{O'Caml} package)                                     \\
 % Fill in if necessary, otherwise leave out.
{\em Subprograms used:} none                                      \\
 %Fill in if necessary, otherwise leave out.
{\em Nature of problem:}\\
 %Describe the nature of the problem here.
 Clebsch-Gordan coefficients are widely used in physics. This program
 has been written as a means to analyze symmetry breaking in the
 context of grand unified theories in particle physics. As an
 example, we computed the singlets appearing in higher-dimensional 
 operators $\bf 27\otimes 27 \otimes 27 \otimes 78$ and $\bf
 27\otimes 27 \otimes 27 \otimes 650$ in an $E_6$-symmetric GUT.  \\ 
{\em Solution method:}\\
 %Describe the method solution here.
 In contrast to very efficient algorithms that also produce tensor
 product decompositions (as far as outer multiplicities /
 Littlewood-Richardson coefficients are concerned) we proceed
 straightforwardly by constructing all the weight states, i.e. the
 Clebsch-Gordan coefficients. This obviously comes at the expense of
 high memory and CPU-time demands. Applying Dynkin arithmetic in
 weight space, the algorithm is an extension of the one for the
 addition of angular momenta in $su(2)\approx A_1$, for reference
 see [1]. Note that, in general, Clebsch-Gordan coefficients are
 basis-dependent and therefore need to be understood with respect to
 the chosen basis. However, singlets appearing in (multiple) tensor
 products are less basis-dependent.  \\ 
{\em Restrictions:}\\
 %Describe any restrictions on the complexity of the problem here.
 Generically, only tensor products of non-degenerate or adjoint
 representations can be computed. However, the irreps appearing
 therein can subsequently be used as new input irreps for further
 tensor product decomposition so in principle there is no restriction
 on the irreps the tensor product is taken of. In practice, though,
 it is by the very nature of the explicit algorithm that input is
 restricted by memory and CPU runtime requirements.  \\
{\em Unusual features:}\\
 %Describe any unusual features of the program/problem here.
 Analytic computation instead of float numerics.  \\
{\em Additional comments:}\\
 %Provide any additional comments here.
 The program can be used in ``notebook style'' using a suitable \texttt{O'Caml}
 toplevel. Alternatively, an \texttt{O'Caml} input file can be compiled which
 effects in processing that is approximately a factor of five
 faster. The latter mode is commendable when large irreps need to be
 constructed. \\
{\em Running time:}\\
 %Give an indication of the typical running time here.
 Varies depending on the input from parts of seconds to weeks for very
 large representations (because of memory
 exhaustion). 

{\em References:}
\begin{refnummer}
\item I. Koh, J. Patera and C. Rousseau, Clebsch-Gordan
   coefficients for $E_6$ and $SO(10)$ unification models;
     G. W. Anderson and T. Blazek, $E_6$
   unification model building I: Clebsch-Gordan coefficients of $\bf
   27 \otimes \bf \overline{27}$.
\end{refnummer}
\end{small}

%%%%%%%%%%%%%%%%%%%%%%%%%%%%%%%%%%%%%%%%%%%%%%%%%%%%%%%%%%%%%%%%

\newpage

\section{Introduction}

We aim at decomposing tensor products of irreducible representations
(irreps) of simple Lie algebras $A$ to $G$. By means of explicitly
constructing weight states via successive lowering operations and
complementing dominant weight spaces we compute Clebsch-Gordan
coefficients. 

This work is organized as follows: in Section \ref{s:theory} we
introduce our notations for Lie algebras and discuss their basic
properties. Then, we develop the machinery that is needed to
explicitly decompose tensor products. In particular, we specify the
basis vectors for which our algorithm yields Clebsch-Gordan
coefficients. Section \ref{s:overview} is an overview of the software
structure whose individual components are described in Section
\ref{s:components}. Installation instructions and test run examples
are given in Sections \ref{s:installation} and \ref{s:test},
respectively. 

%%%%%

\section{Theoretical background}
\label{s:theory}

\subsection{Lie algebras: basic facts and representation theory}

In this subsection we summarize the basic facts about the
representation theory of Lie algebras to make the paper self-contained
and to fix our notation. Most properties discussed here can be found
in the literature, e.g.~\cite{Fuchs:1997jv,Georgi:1982jb,Slansky:1981yr,Cornwell}.

A Lie algebra is usually defined via the fundamental Lie bracket,
\begin{equation}
 \left[ T^a, T^b \right] = i f^{abc} T^c \quad ,
\end{equation}
where the Lie algebra is completely determined by knowledge of its
structure constants, $f^{abc}$. The Lie bracket fulfils the Jacobi
identity, 
\begin{equation}
 \left[T^a, \left[ T^b , T^c \right] \right] + 
 \left[T^b, \left[ T^c , T^a \right] \right] + 
 \left[T^c, \left[ T^a , T^b \right] \right] = 0 \quad .
\end{equation}

A Lie algebra is called simple if it is neither Abelian, i.e. if
$\left[ \mathfrak{g}, \mathfrak{g} \right] \neq 0$, nor has any
non-trivial ideal, i.e. a subalgebra $\mathfrak{h} \subset
\mathfrak{g}$ with $\left[ \mathfrak{h} , \mathfrak{g} \right] =
\mathfrak{h}$ other than the zero element and the algebra
$\mathfrak{g}$ itself. The simple Lie algebras can be classified by
simple geometric methods~\cite{sophus,Dynkin:1957um}, and can be
grouped into the four infinite families of rank-$n$ algebras, namely
the special unitary algebras $A_n \equiv \mathfrak{su}(n+1)$, the
special orthogonal algebras of odd dimension, $B_n \equiv
\mathfrak{so}(2n+1)$, the symplectic algebras $C_n \equiv
\mathfrak{sp}(2n)$, as well as the special orthogonal algebras of even
dimension, $D_n \equiv \mathfrak{so}(2n)$. Furthermore, there are
five so-called exceptional Lie algebras, specified by their
corresponding ranks, $E_6$, $E_7$, $E_8$, $F_4$, and $G_2$.  

A representation (rep) of a Lie algebra is a homomorphism from the set
of generators to a corresponding set of linear mappings, $R_V$, over a 
vector space, $V$, denoted in a compact way: $(V, R_V)$. The
dimension of $V$ is called the dimension of the representation. 

A maximal subset of mutually commuting Hermitian generators, $H_i$,
$i= 1, \ldots,n$ is called a Cartan subalgebra. Its dimension is the
rank of the Lie algebra. Its elements, the Cartan generators, fulfil  
\begin{equation}
  H_i = H_i^\dagger \qquad  \left[ H_i , H_j \right] = 0.
\end{equation}
Hence, they can be simultaneously diagonalized. As they form a linear
space, we can choose a basis where $\tr \left[H_i H_j\right] = k_D
\delta_{ij}$, where the Dynkin index $k_D$ depends on the
representation and the normalization of the generators. Any state in a
representation $D$ can be (up to possible degeneracies from other
symmetries) uniquely labelled by the eigenvalues of the corresponding
Cartan generators:
\begin{equation}
  H_i \ket{\mu, D} = \mu_i \ket{\mu, D} . 
\end{equation}
The eigenvalues $\mu_i$ are called weights, they can be grouped
together in a vector of the dimension of the rank, called a weight
vector. A scalar product on weight vectors $\alpha \cdot \mu \equiv
\alpha_i \mu_i$ is defined via index summation over $i=1,\ldots,
\text{rank}(\mathfrak{g})$. 

A special representation is the adjoint representation, where the
generators act by means of the Lie bracket on themselves, $T^a
\ket{T^b} := \ket{ \left[ T^a, T^b \right]}$. Hence, the adjoint
representation has the dimension of the rank of the algebra. Most of
the facts about representations, weights etc. can be found in
\cite{Georgi:1982jb,Slansky:1981yr}. The weights of the adjoint representation
are called roots. The non-Cartan generators, labelled by their root
vectors $\alpha$ as $E_\alpha$, come in pairs of adjoints
to each other with $E_{-\alpha} = E_\alpha^\dagger$. These form
$SU(2)$ subalgebras of the Lie algebra under consideration with
$E_\alpha$ and $E_{-\alpha}$ being the corresponding raising and
lowering operators:  
\begin{equation}
  \left[ E_\alpha , E_{-\alpha} \right] = \alpha \cdot H \quad .
\end{equation}
This choice of operators is called the Cartan-Weyl basis.
As the finite-dimen\-sio\-nal unitary representations of $SU(2)$ are
well-known, all reps of all Lie algebras can be constructed. Let $p$
be the maximum number a raising operator can be applied to an
arbitrary state $\ket{\mu, D}$ of a rep $D$, $q$ the maximum number
one can descend, then $SU(2)$ algebra allows to relate the differences
of the numbers $(p-q)$ and $(p'-q')$ for two different roots
corresponding to two different raising operators to an angle between
the roots, and it can be shown that this angle can only be 90, 120,
135, or 150 degrees. 

To specify a highest weight of a rep (like $m = +j$ for spin $j$ in
quantum mechanics) one introduces an order on the space of weights by
defining a weight to be positive if its first non-vanishing component
is positive. With this help one can find a basis in the root space of
a Lie algebra by taking the so-called simple roots. These are those
positive roots which cannot be written as sums of other positive
roots. There are as many simple roots as the rank of the algebra.

This positivity definition allows immediately for the ordering, namely
two weights are ordered, $\mu > \nu$ if $\mu - \nu$ is positive. In
the adjoint representation, the positive roots correspond to raising
operators and the negative roots to lowering operators. The highest
weight by definition cannot be raised by any raising operator. 

There is a diagrammatic notion of the system of simple roots, the
Dynkin diagrams. Here, simple roots are denoted by circles, which are
unconnected if the simple roots are orthogonal to each other, and
connected simply, doubly or triply, if the enclosed angle between the
two roots is 120, 135, or 150 degrees, respectively. Our conventions
for the labelling of the roots are shown in Fig.~\ref{fig:dynkin} in
Appendix \ref{sec:dynkin}. Actually, there are extended Dynkin
diagrams shown for all simple Lie algebras, augmented by the lowest
root which is important for the determination of the maximal
subalgebras. Shorter roots (by a factor $\sqrt{2}$ for double links, a
factor $\sqrt{3}$ for triple lines) are denoted by filled circles.

The Cartan matrix allows to calculate the difference between lowering
and raising possibilities for states corresponding to a positive root, 
$\phi = \sum_j k_j \alpha^j$:
\begin{equation}
  q^i - p^i = \sum_j k_j A_{ji} \qquad \text{with} \qquad
  \boxed{ A_{ji} \equiv \frac{2 \alpha^j \cdot
  \alpha^i}{{\alpha^i}^2}} \quad .
\end{equation}
Note that this is the definition in \cite{Georgi:1982jb,Slansky:1981yr}, but
the transpose of \cite{Cornwell}. The $j$th row of the Cartan matrix
consists of the $q^i - p^i$ values of the simple root $\alpha^j$. 
The Cartan matrices in our notation can be found in Appendix
\ref{sec:cartan}. 

By this means, it is straightforward to construct all possible reps of
all simple Lie algebras, and there are many examples in the literature
\cite{Georgi:1982jb,Slansky:1981yr}. 

Any irrep is uniquely determined by its so-called Dynkin coefficients,
$\ell^j$, which are defined via
\begin{equation}
  \frac{2 \alpha^j \cdot \mu}{{\alpha^j}^2} = \ell^j \quad . 
\end{equation}
Every set of $\ell^j$ gives a $\mu$ which is the highest weight of
one irrep. Hence, we choose as input a sequence of
$\text{rank}(\mathfrak{g})$ non-negative integers to specify an
irrep. Defining as fundamental weights those weight vectors for which
$2\alpha^j \cdot \mu^k / {\alpha^j}^2 = \delta_{jk}$, every highest
weight can then be uniquely written as $\mu =
\sum_{j=1}^{\text{rank}(\mathfrak{g})} \ell^j \mu^j$. The reps
corresponding to those fundamental weights are the fundamental reps,
again their number is $\text{rank}(\mathfrak{g})$. 

As an example we show the construction of all states of the
fundamental irreps {\bf 27} of $E_6$. We start from the fundamental
weight 100000 as highest weight, and descend with the help of the
Cartan matrix from it until we reach the state with the lowest weight
at the bottom. What is shown in addition is the Dynkin coefficient of
the lowest root. 

\begin{figure}
  \begin{center}
  \includegraphics[width=.8\textwidth]{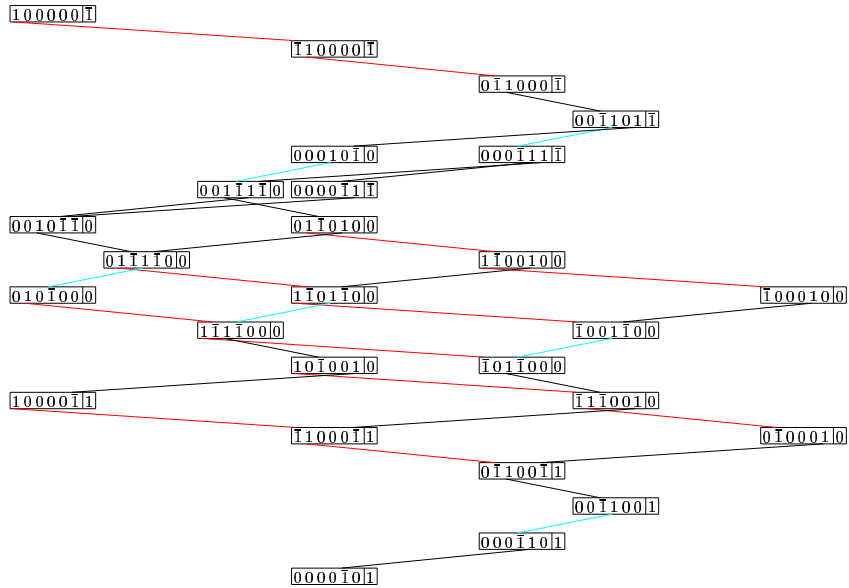}
  \end{center}
  \caption{Example for the construction of a rep of a Lie algebra, the
    fundamental {\bf 27} of $E_6$, with Dynkin coefficients $100000$.}
  \label{fig:27ofe6}
\end{figure}
The program \texttt{CleGo} has a subprogram which is independently steerable
by the user called \texttt{lie.\{opt|bin\}}. From the command line,
the user specifies the Lie algebra under consideration as well as the
irrep to be constructed. E.g. to produce the adjoint rep of $SU(3)$
one uses \texttt{./lie.opt -su 3 -rep 11} which yields
\begin{scriptsize}
\begin{verbatim}
                            Lie algebra   :   SU(3)
                            ==================================
                            Highest weight:   (1,1)
                            Dim. of irrep :   8
                            ==================================
                            1, Lev:0, Deg:1  (1,1),-2  (0,0)
                            2, Lev:1, Deg:1  (2,-1),-1  (0,1)
                            3, Lev:1, Deg:1  (-1,2),-1  (1,0)
                            4, Lev:2, Deg:2  (0,0),0  (1,1)
                            6, Lev:3, Deg:1  (1,-2),1  (1,2)
                            7, Lev:3, Deg:1  (-2,1),1  (2,1)
                            8, Lev:4, Deg:1  (-1,-1),2  (2,2)
\end{verbatim}
\end{scriptsize}

To reproduce the example from Fig. \ref{fig:27ofe6} 
of the irrep {\bf 27} of $E_6$ one
types in \texttt{./lie.opt -e6 -rep 100000}.  This produces:
\newpage
\begin{scriptsize}
\begin{verbatim}
                            Lie algebra   :   E6
                            ==================================
                            Highest weight:   (1,0,0,0,0,0)
                            Dim. of irrep :   27
                            ==================================
                            1, Lev:0, Deg:1  (1,0,0,0,0,0),-1  (0,0,0,0,0,0)
                            2, Lev:1, Deg:1  (-1,1,0,0,0,0),-1  (1,0,0,0,0,0)
                            3, Lev:2, Deg:1  (0,-1,1,0,0,0),-1  (1,1,0,0,0,0)
                            4, Lev:3, Deg:1  (0,0,-1,1,0,1),-1  (1,1,1,0,0,0)
                            5, Lev:4, Deg:1  (0,0,0,1,0,-1),0  (1,1,1,0,0,1)
                            6, Lev:4, Deg:1  (0,0,0,-1,1,1),-1  (1,1,1,1,0,0)
                            7, Lev:5, Deg:1  (0,0,1,-1,1,-1),0  (1,1,1,1,0,1)
                            8, Lev:5, Deg:1  (0,0,0,0,-1,1),-1  (1,1,1,1,1,0)
                            9, Lev:6, Deg:1  (0,0,1,0,-1,-1),0  (1,1,1,1,1,1)
                            10, Lev:6, Deg:1  (0,1,-1,0,1,0),0  (1,1,2,1,0,1)
                            11, Lev:7, Deg:1  (0,1,-1,1,-1,0),0  (1,1,2,1,1,1)
                            12, Lev:7, Deg:1  (1,-1,0,0,1,0),0  (1,2,2,1,0,1)
                            13, Lev:8, Deg:1  (0,1,0,-1,0,0),0  (1,1,2,2,1,1)
                            14, Lev:8, Deg:1  (1,-1,0,1,-1,0),0  (1,2,2,1,1,1)
                            15, Lev:8, Deg:1  (-1,0,0,0,1,0),0  (2,2,2,1,0,1)
                            16, Lev:9, Deg:1  (1,-1,1,-1,0,0),0  (1,2,2,2,1,1)
                            17, Lev:9, Deg:1  (-1,0,0,1,-1,0),0  (2,2,2,1,1,1)
                            18, Lev:10, Deg:1  (1,0,-1,0,0,1),0  (1,2,3,2,1,1)
                            19, Lev:10, Deg:1  (-1,0,1,-1,0,0),0  (2,2,2,2,1,1)
                            20, Lev:11, Deg:1  (1,0,0,0,0,-1),1  (1,2,3,2,1,2)
                            21, Lev:11, Deg:1  (-1,1,-1,0,0,1),0  (2,2,3,2,1,1)
                            22, Lev:12, Deg:1  (-1,1,0,0,0,-1),1  (2,2,3,2,1,2)
                            23, Lev:12, Deg:1  (0,-1,0,0,0,1),0  (2,3,3,2,1,1)
                            24, Lev:13, Deg:1  (0,-1,1,0,0,-1),1  (2,3,3,2,1,2)
                            25, Lev:14, Deg:1  (0,0,-1,1,0,0),1  (2,3,4,2,1,2)
                            26, Lev:15, Deg:1  (0,0,0,-1,1,0),1  (2,3,4,3,1,2)
                            27, Lev:16, Deg:1  (0,0,0,0,-1,0),1  (2,3,4,3,2,2)    
\end{verbatim}
\end{scriptsize}
The program repeats the Lie algebra, the highest weight chosen,
calculates the dimension of the irrep, and then constructs it. 
In each line, the entries are the number of the state, the level below
the highest weight (with the highest weight having level zero), the
level of degeneracy (here there is only the trivial degeneracy of the
$\text{rank}(\mathfrak{g})$ zero weights of the adjoint). The two
rightmost columns show the Dynkin coefficients of the corresponding
weight with the Dynkin coefficient of the lowest root split by a
comma, and the number of applications of the corresponding lowering
operators from the highest weight. E.g. state number 4 of the {\bf 27}
of $E_6$ is generated by applying the raising operators corresponding
to the simple roots $\alpha^{1,2,3}$ exactly once. 

While the construction of the weights of an irrep is just simple
addition of rows of the corresponding Cartan matrix to the highest
weight, the determination of the dimension of an irrep can be done by
means of the system of the positive roots of the Lie algebra
$\mathfrak{g}$, $\Delta_+$. The roots of a Lie algebra can be uniquely
decomposed (up to Weyl symmetries of the root lattice) into the zero
roots, the positive and the negative roots. Our conventions for the
positive roots can be found in Appendix \ref{seq:positiveroots}, again
in the Dynkin basis. In the following discussion we denote the highest
weight of an irrep in the Dynkin basis by $\Lambda =
\sum_{j=1}^{\text{rank}(\mathfrak{g})} \ell^j \mu^j$. 

The dimension of an irrep $R$ can be derived from characters and is given
by Weyl's formula \cite{weyl}:
\begin{equation}
  \label{eq:weyl}
  \dim(R) = \prod_{\alpha \in \Delta_+} \frac{(\Lambda + \delta)\cdot
    \alpha}{\delta \cdot \alpha} = \prod_{\alpha\in\Delta_+} \left[
    \frac{\sum_{j=1}^{\text{rank}(\mathfrak{g})} n_j k_j^\alpha
      \omega_j}{\sum_{j=1}^{\text{rank}(\mathfrak{g})} k_j^\alpha
      \omega_j} + 1 \right]
\end{equation}
where $\delta = \frac12 \sum_{\alpha\in\Delta_+} \alpha$. $k_j^\alpha$
are the coefficients of the root $\alpha$ with respect to the simple
roots, $\alpha = \sum_{j=1}^{\text{rank}(\mathfrak{g})} k_j^\alpha
\alpha^j$, $n_j$ are the Dynkin labels of the highest weight, $\Lambda
= \sum_{j=1}^{\text{rank}(\mathfrak{g})} n_j \mu^j$, and $\omega_j$ is
the weight of the simple root $\alpha^j$ in the Dynkin diagram. The
weights of the roots are their squares, they are one for all roots in
Lie algebras with only simple links, one for the shorter and two for
the longer roots in $SO(2n+1)$, $Sp(2n)$, and $F_4$, and one for the
shorter, three for the longer root in $G_2$. 

The degree of degeneracy for a specific state in an irrep can be
calculated by the Freudenthal recursion formula \cite{freudenthal}
recursively from the degeneracies of all states of higher levels than
the corresponding state. The formula goes as follows: for an irrep $R$
with highest weight $\Lambda$ the multiplicity $m(\lambda)$ of a
possible weight $\lambda = \Lambda -
\sum_{j=1}^{\text{rank}(\mathfrak{g})} q_j \alpha^j$ with $q_1, q_2,
\ldots, q_{\text{rank}(\mathfrak{g})}$ all non-negative integers is
given by: 
\begin{equation}
  \label{eq:freudenthalabstract}
  \left[ (\Lambda + \delta)\cdot (\Lambda + \delta) - 
    (\lambda + \delta) \cdot (\lambda + \delta) \right] m(\lambda) = 
  2 \sum_{\alpha\in\Delta_+} \sum_k m(\lambda + k \alpha) ((\lambda +
  k \alpha) \cdot \alpha) 
\end{equation}
The second sum on the right hand side extends over all those values of
$k$ for which $\lambda + k \alpha$ whose level is less than that of
$\lambda$. $\delta = \frac12 \sum_{\alpha\in\Delta_+} \alpha$ as in
the Weyl formula. The Freudenthal formula can be recast in a more
usable form which has been used in the implementation in our program: 
\begin{multline}
  \label{eq:freudenthal}
  \left\{  \sum_{j=1}^{\text{rank}(\mathfrak{g})} q_j \omega_j \left[ 
      (n_j + 1) - \frac12 \sum_{i=1}^{\text{rank}(\mathfrak{g})} q_i
      A_{ji} \right] \right\} m(\lambda)  \\
  = \sum_{\alpha\in\Delta_+} \sum_k m(\lambda+k\alpha) \left\{ 
    \sum_{j=1}^{\text{rank}(\mathfrak{g})} k_j^\alpha \omega_j \left[
      n_j + \sum_{i=1}^{\text{rank}(\mathfrak{g})} (kk_i^\alpha - q_i)
    \right] \right\} 
\end{multline}
Here, $\Lambda = \sum_{j=1}^{\text{rank}(\mathfrak{g})} n_j \mu^j$,
$\lambda = \Lambda - \sum_{j=1}^{\text{rank}(\mathfrak{g})} q_j
\alpha^j$, $\alpha = \sum_{j=1}^{\text{rank}(\mathfrak{g})}
k_j^\alpha \alpha^j$, and $\omega_i$ is the weight of the simple root
$\alpha^i$ in the Dynkin diagram. The second sum is defined in the
same way as the one in (\ref{eq:freudenthalabstract}). The highest
weight has $m(\Lambda) = 1$ which allows to obtain the degeneracies or
multiplicities first for level 1, then for level 2, and so on. For
example, every level-1 weight has the form $\lambda = \Lambda -
\alpha^j$, where $\alpha^j$ is a simple root. The only non-vanishing
term on the right-hand side of \eqref{eq:freudenthal} is for $\alpha
= \alpha^j$ and $k=1$, which is $2 m(\Lambda) \Lambda \cdot \alpha^j =  
\Lambda \cdot \alpha^j$. In the next step, the degeneracies of the
level-2 weights are determined from those of level 0 and 1, and so on.

Regarding the determination of the Dynkin coefficient of the lowest
root, $\alpha^0$, we have to determine first the linear combination 
with respect to the simple roots. These are (in our conventions):
\begin{subequations}
  \label{eq:lowestroot}
\begin{align}
  A_n: \qquad \alpha^0 &=\; - \sum_{i=1}^n \alpha^i \\ 
  B_n: \qquad \alpha^0 &=\; - \alpha^1 - 2 \sum_{i=2}^n \alpha^i \\ 
  C_n: \qquad \alpha^0 &=\; - 2 \sum_{i=1}^{n-1} \alpha^i - \alpha^n \\ 
  D_n: \qquad \alpha^0 &=\; - \alpha^1 - 2 \sum_{i=2}^{n-3} \alpha^i 
  - \alpha^{n-1} - \alpha^n \\ 
  E_6: \qquad \alpha^0 &=\; - \alpha^1 - 2 \alpha^2 - 3 \alpha^3 - 2
  \alpha^4 - \alpha^5 - 2 \alpha^6 \\
  E_7: \qquad \alpha^0 &=\; - 2 \alpha^1 - 3 \alpha^2 - 4 \alpha^3 - 3
  \alpha^4 - 2 \alpha^5 - \alpha^6 - 2 \alpha^7 \\ 
  E_8: \qquad \alpha^0 &=\; - 2 \alpha^1 - 4 \alpha^2 - 6 \alpha^3 - 5
  \alpha^4 - 4 \alpha^5 - 3 \alpha^6 - 2 \alpha^7 - 3 \alpha^8 \\ 
  F_4: \qquad \alpha^0 &=\; - 2 \alpha^1 - 4 \alpha^2 - 3 \alpha^3 - 2
  \alpha^4 \\
  G_2: \qquad \alpha^0 &=\; -3 \alpha^1 - 2 \alpha^2
\end{align}
\end{subequations}

To calculate the Dynkin coefficients according to the formula
\begin{equation}
  \ell^0 = \frac{2 \alpha^0 \cdot \mu}{{\alpha^0}^2} \quad ,
\end{equation}
the formulae (\ref{eq:lowestroot}) are the same for the Dynkin
coefficients for the series $A_n$ and $D_n$ as well as for
$E_{6,7,8}$. For $F_4$, $\alpha^1$ and $\alpha^2$ are shorter by a
factor $\sqrt{2}$ as the other three, leaving us with the relation 
$\ell^0 = - \ell^1 - 2 \ell^2 - 3 \ell^3 - 2 \ell^4$. In the case of
$G_2$, $\alpha^0$ and $\alpha^2$ are longer by a factor $\sqrt{3}$ as
$\alpha^1$, which yields $\ell^0 = - \ell^1 - 2 \ell^2$. For the two
infinite series $B_n$ the last coefficient $\ell^n$ has to be rescaled
by two, while for $C_n$ all coefficients except for the last one have to
be rescaled by two. In general, $\alpha^0$ is always among the longer
roots.  

%%%%%

\subsection{Tensor product representation}

The definition of the tensor product of representations of a Lie
algebra is as follows: Let $\mathfrak{g}$ be a Lie algebra and $(V,
R_V)$, $(W, R_W)$ representations of $\mathfrak{g}$. We then construct 
another representation, the so-called tensor product representation,
$(V\otimes W, R_V  \otimes R_W)$ via 
\begin{eqnarray}
 \label{def_tp_rep}
 (R_V \otimes R_W) (x): V\otimes W \rightarrow V\otimes W
 \nonumber\\
 \left((R_V\otimes R_W)(x) \right)(v\otimes w) := (R_V(x)v)\otimes w
 + v \otimes (R_W(x) w)
\end{eqnarray}
for all $x\in\mathfrak{g}$. If the original representations are
finite-dimensional and irreducible, the tensor product representation
is also finite-dimensional which in the case of (semi)simple algebras
implies that it is completely reducible  
$$V\otimes W \cong \bigoplus_{i=1}^n U_i, \qquad R_V\otimes R_W \cong
\bigoplus_{i=1}^n R_{U_i} $$ 
with irreducible representations $(U_i, R_{U_i})$. 

The tensor product is commutative up to isomorphism, i.e.\ there is an
isomorphism $f: V\otimes W \rightarrow W\otimes V$ such that
$$R_{W\otimes V}(x)\circ f = f \circ R_{V\otimes W}(x), \qquad\forall
x\in\mathfrak{g}.$$  
Also, multiple tensor products of representations are associative up
to isomorphism which again means that there is an isomorphism $$f:
U\otimes(V\otimes W) \rightarrow (U\otimes V)\otimes W$$ that
intertwines representations $(U\otimes(V\otimes
W),R_U\otimes(R_V\otimes R_W))$ and $((U\otimes V)\otimes W,
(R_U\otimes R_V)\otimes R_W)$ of $\mathfrak{g}$. 

In particular, for a simple Lie algebra and finite-dimensional
representations, we have the decomposition of the tensor product
module 
\begin{equation}
 \label{cg_decomp}
 V_{\Lambda} \otimes V_{\Lambda'} \cong \bigoplus_i
 \mathcal{L}_{\Lambda\Lambda'}^{\Lambda_i} V_{\Lambda_i}
\end{equation}
where modules $V_{\Lambda}$ are uniquely specified by their
highest weight $\Lambda$, etc.. The number of times an irrep
appears on the right-hand side of \eqref{cg_decomp} is given by
$\mathcal{L}_{\Lambda\Lambda'}^{\Lambda_i}$ which are non-negative
integers to be denoted tensor product multiplicities or
Littlewood-Richardson coefficients of $\mathfrak{g}$.

The Cartan-Weyl basis of the Lie algebra is translated to operators in
the respective representations by means of the algebra homomorphism of
the representations. Let its Cartan generators $H_i$ be represented as
$(H_{V_{\Lambda}})_i$ and $(H_{V_{\Lambda'}})_i$ on $V_{\Lambda}$ and
$V_{\Lambda'}$ respectively, the definition of the tensor product
representation in \eqref{def_tp_rep} shows that the Cartan generators
acting on the product space read
\begin{equation}
 \label{tp_cartan_gens}
 H_i = (H_{V_{\Lambda}})_i \otimes 1 + 1 \otimes (H_{V_{\Lambda'}})_i.
\end{equation}
Likewise, raising and lowering operators
$(E_{V_{\Lambda}})_{\pm\alpha}$ and $(E_{V_{\Lambda'}})_{\pm\alpha}$
are given by
\begin{equation}
 \label{tp_cartan_low_rais}
 E_{\pm\alpha} = (E_V)_{\pm\alpha} \otimes 1 + 1 \otimes
 (E_W)_{\pm\alpha}
\end{equation}
where $\alpha$ denote the roots of the Lie algebra $\mathfrak{g}$. One 
can easily check that the generators of the product representation
again form a Cartan-Weyl basis. In particular, we have  
\begin{eqnarray}
 [H_i,H_j] & = & 0, \nonumber\\
 {[H_i,E_{\pm\alpha}]} & = & \pm\alpha_i E_{\pm\alpha}.
\end{eqnarray}
It is therefore again appropriate to discuss the tensor product
representation in terms of weights and roots, that is, in Dynkin
weight space. From \eqref{tp_cartan_gens}, it is clear that all
weights of the tensor product representation are obtained by adding
weights of the respective irreps the tensor product is taken of. The 
multiplicities of these weights $\lambda$ is
$$\text{mult}_{V_{\Lambda} \otimes V_{\Lambda'}} (\lambda) =
\sum_{\mu,\mu' \text{ with } \mu+\mu'=\lambda}
\text{mult}_{V_{\Lambda}}(\mu) \cdot
\text{mult}_{V_{\Lambda'}}(\mu').$$ 
It then follows that the highest weight in $V_{\Lambda}\otimes
V_{\Lambda'}$ is given by the sum of the highest weights of
$V_{\Lambda}$ and $V_{\Lambda'}$, i.e.\ $\Lambda+\Lambda'$, and
appears with multiplicity one. 

%%%%%%

\subsection{Explicit lowering}
\label{explicit_lowering}
Before describing the decomposition algorithm we need to discuss in
some detail how the Cartan-Weyl generators act on representation
space. This discussion is close to the one in \cite{Anderson:2001sd}. Given an
irrep of a simple Lie algebra we choose basis states in the weight
subspaces that, in general, are degenerate. In what follows we use the
following basis states
\begin{eqnarray*}
 | (w+\alpha)_{\Gamma}\rangle, & \, & \text{where }
 \Gamma=1,\ldots,D_{w+\alpha},\\ 
 | w_c \rangle, & \, & \text{where }
 c=1,\ldots,D_{w},\\
 | (w-\alpha)_C \rangle, & \, & \text{where }
 C=1,\ldots,D_{w-\alpha}, 
\end{eqnarray*}
where $D_w$ denotes the degeneracy of the subspace of weight
$w$. Those basis states are in general non-orthogonal. Without loss of
generality, we want them to be always normalized to unity, i.e.\   
$$\langle w_c | w_c \rangle = 1, \qquad \forall c.$$
For every weight $w$ in the representation, let $M^{w}$ be a matrix
with $M_{ab}^{w} = \langle w_a | w_b \rangle$ and $G^{w}$ the matrix
with $G_{ab}^{w}=((M^{w})^{-1})_{ab}$, then the identity operator
in the degenerate subspace is given by
\begin{equation}
 \label{expl_low_id}
 I^{(w)} = \sum_{a,b} G_{ab}^{w} | w_a\rangle\langle w_b |.
\end{equation}
It helps to consider states 
\begin{eqnarray*}
 |\hat{w}_b \rangle & = & \sum_a | w_a\rangle G_{ab}^{w} \\
 \langle \hat{w}_a| & = & \sum_b G_{ab}^{w} \langle w_b|,
\end{eqnarray*}
which satisfy
\begin{equation}
 \label{expl_low_kron}
 \langle w_c|\hat{w}_a\rangle = \langle \hat{w}_a | w_c \rangle =
 \delta_{ac}.
\end{equation}
Generically, raising or lowering of a degenerate weight state is a
linear combination of degenerate states of the adjacent weights:
\begin{eqnarray}
 \label{expl_low_def}
 E_{\alpha_i} |w_c\rangle & = & \sum_{\Gamma} N_{\alpha_i, w_c
   \rightarrow (w+\alpha_i)_{\Gamma}} |(w+\alpha_i)_{\Gamma}\rangle
 \nonumber\\
 E_{-\alpha_i} |w_c\rangle & = & \sum_{C} N_{-\alpha_i, w_c
   \rightarrow (w-\alpha_i)_{C}} |(w-\alpha_i)_{C}\rangle
\end{eqnarray}
Using \eqref{expl_low_kron} we can single out the lowering coefficient
\begin{equation}
 \label{expl_low_N}
 N_{-\alpha_i, w_a \rightarrow (w-\alpha_i)_{A}} = \sum_B
 G_{AB}^{w-\alpha_i} \langle (w-\alpha_i)_B | E_{-\alpha_i} | w_a \rangle.
\end{equation}
From $E_{\alpha} = E_{-\alpha}^{\dag}$ and the defining relation in
\eqref{expl_low_def} we obtain
\begin{eqnarray*}
 & \sum_{A,B} & \!\!N_{-\alpha, w_a\rightarrow (w-\alpha)_A}
 N^*_{-\alpha,w_b\rightarrow(w-\alpha)_B}
 \langle(w-\alpha)_B|(w-\alpha)_A\rangle\\   
 & = & \!\! \langle w_b |E_{\alpha} E_{-\alpha} | w_a \rangle
 = \langle w_b|[E_{\alpha},E_{-\alpha}]+E_{-\alpha} E_{\alpha} |
 w_a\rangle \\ 
 & = & \!\!\langle w_b | w_a \rangle \langle\alpha,w\rangle + 
 \sum_{\Gamma,\Delta} G_{\Gamma\Delta}^{w+\alpha} \langle w_b| E_{-\alpha}
 |(w+\alpha)_{\Gamma}\rangle\langle(w+\alpha)_{\Delta}| E_{\alpha} |
 w_a\rangle \\ 
 & = & \!\! \langle w_b | w_a \rangle \langle\alpha,w\rangle + \\
 && \qquad\quad
 \sum_{\Gamma,\Delta,c,d} G_{\Gamma\Delta}^{w+\alpha} N_{-\alpha,
   (w+\alpha)_{\Gamma}\rightarrow w_c} \langle w_b | w_c \rangle
 N^*_{-\alpha,(w+\alpha)_{\Delta}\rightarrow w_d} \langle w_d|w_a\rangle
\end{eqnarray*}
where we made use of $[E_{\alpha}, E_{-\alpha}] =
\langle\alpha,H\rangle$ and the identity operator from
\eqref{expl_low_id} in the next to last step, as well as
\eqref{expl_low_N} in the last step. Note that in the Dynkin basis,
$\langle\alpha_i,w\rangle$ denotes the $i$-th Dynkin coordinate of $w$,
i.e.\ $w_i$. 

Hence, setting $a=b$ we obtain
\begin{eqnarray}
 \label{expl_low_rec}
 &\sum_{A,B}&\left((G^{w-\alpha})^{-1}\right)_{AB} N_{-\alpha,
   w_a\rightarrow(w-\alpha)_A} N^*_{-\alpha,
   w_a\rightarrow(w-\alpha)_B} \nonumber\\
 &=&\langle\alpha,w\rangle + \nonumber\\ 
 && \qquad
 \sum_{\Gamma,\Delta, c,d}
 G_{\Gamma\Delta}^{w+\alpha} \left((G^w)^{-1}\right)_{ac}
 \left((G^w)^{-1}\right)_{ad} N_{-\alpha, (w+\alpha)_{\Gamma}
   \rightarrow w_c} N^*_{-\alpha, (w+\alpha)_{\Delta} \rightarrow
   w_d},  \nonumber\\
\end{eqnarray}
which is close to a recursion relation for the lowering
normalizations: Given that one knows how to lower states of weight
$w+\alpha$ and one also has the scalar products of the basis states of
both weight $w$ and $w+\alpha$, the right-hand side of
\eqref{expl_low_rec} is completely determined. We see that
normalizations for lowering states of weight $w$ and scalar products
of states of weight $w-\alpha$ need to be chosen consistently, that is,
in such a way that \eqref{expl_low_rec} holds. 

There are two special cases in which the consistency equation in
\eqref{expl_low_rec} simplifies considerably. First, in the case
where the irrep is non-degenerate, we have by construction 
$$M_{ab}^w=\langle w_a | w_b \rangle = 1, \quad G_{ab}^w=
\left((M^w)^{-1}\right)_{ab}=1, \qquad a,b=1,\,\forall w$$ 
which simplifies \eqref{expl_low_rec} to
\begin{equation}
 \label{expl_low_non_deg_irrep}
 | N_{-\alpha, w\rightarrow (w-\alpha)} |^2 = \langle\alpha,w\rangle
 +  |N_{-\alpha, (w+\alpha)\rightarrow w}|^2. 
\end{equation}
Thus, for non-degenerate irreps lowering normalizations are
recursively determined up to the choice of phase factors. It is
convenient to always take lowering phase factors to be unity. If the
right-hand side turns out to be zero or negative, the state
$|w-\alpha\rangle$ does not exist.

Secondly, we consider the case where the irrep is the adjoint
representation which has a degenerate zero weight space but
is otherwise non-degenerate. Starting with lowering the highest
weight, normalization constants are the same as in the non-degenerate
case until we are to lower the simple roots. For a simple
root\footnote{Upper indices label distinct simple roots in order to
 distinguish them from lower indices that label degeneracies.}  
$w=\alpha^i$ lowering can only occur via $\alpha=\alpha^i$ and the
consistency equation becomes  
\begin{equation*}
 \sum_{A,B=1}^{\text{rank}} \left(M^0 \right)_{AB} N_{-\alpha^i,
 \alpha^i\rightarrow 0_A} N^*_{-\alpha^i, \alpha^i\rightarrow 0_B} =
 \langle\alpha^i,\alpha^i\rangle + |N_{-\alpha^i, 2\alpha^i
   \rightarrow \alpha^i}|^2. 
\end{equation*}
Now the basis in the subspace of weight zero is chosen in such a way 
that basis states therein result from lowering simple roots and are
normalized to unity, i.e. 
\begin{equation}
 \label{expl_low_n_simpl_r}
 N_{-\alpha^i, \alpha^i\rightarrow 0_A} = \left \{
 \begin{matrix}
   N_{-\alpha^i, \alpha^i\rightarrow 0_i}, & \text{if }
   A=i   \\ 
   0, & \mbox{else}
 \end{matrix}\right .
\end{equation}
and obtain
\begin{eqnarray}
 \label{expl_low_n_simpl_r2} 
 &(M^0)_{ii} & \!\! |N_{-\alpha^i, \alpha^i\rightarrow 0_i}|^2 =
 \langle\alpha^i,\alpha^i\rangle + \underbrace{|N_{-\alpha^i, 2\alpha^i
   \rightarrow \alpha^i}|^2}_{=0}
 \nonumber\\
 & \Rightarrow & \!\! |N_{-\alpha^i, \alpha^i\rightarrow 0_i}|^2 = 2.
\end{eqnarray}
Note that in \eqref{expl_low_n_simpl_r2} $|N_{-\alpha^i, 2\alpha^i
 \rightarrow \alpha^i}|^2$ vanishes because $2\alpha^i$ is never a
root. Now, if $w=0_a$ and $\alpha=\alpha^b$, we have  
\begin{eqnarray}
 \label{expl_low_lowering_zero}
 |N_{-\alpha^b, 0_a\rightarrow -\alpha^b}|^2
 &=&\underbrace{\langle\alpha^b,0\rangle}_{=0} + \sum_{c,d} (M^0)_{ac}
 (M^0)_{ad} N_{-\alpha^b, \alpha^b\rightarrow 0_c}
 N^*_{-\alpha^b,\alpha^b\rightarrow 0_d} \nonumber\\
 &=&(M^0)_{ab} (M^0)_{ab} \, |N_{-\alpha^b,
   \alpha^b\rightarrow 0_b}|^2 \nonumber\\
 &=&2 \langle 0_a|0_b\rangle^2 = \frac{A_{ab} A_{ba}}{2}
\end{eqnarray}
where \eqref{expl_low_n_simpl_r} and \eqref{expl_low_n_simpl_r2} have
been used and the last step is due to 
\begin{equation}
 \label{expl_low_scp_zeros_adjoint}
 \langle 0_a|0_b\rangle = 
 \frac{\sqrt{A_{ab}A_{ba}}}{2}
\end{equation}
with the Cartan matrix $A$. Relation
\eqref{expl_low_scp_zeros_adjoint} will be shown in
\ref{sec:scp}. Now let $w=-\alpha^a$ and $\alpha=\alpha^b$, then the
condition in \eqref{expl_low_rec} reads  
\begin{multline*}
 |N_{-\alpha^b, -\alpha^a \rightarrow - \alpha^a-\alpha^b} |^2 =
 \\
 \langle\alpha^b,-\alpha^a\rangle + \sum_{\Gamma,\Delta}
 (G^{\alpha^b-\alpha^a})_{\Gamma\Delta} N_{-\alpha^b,
   (\alpha^b-\alpha^a)_{\Gamma} \rightarrow -\alpha^a} N^*_{-\alpha^b,
   (\alpha^b-\alpha^a)_{\Delta} \rightarrow -\alpha^a}
\end{multline*}
which for $a=b$ becomes (using \eqref{expl_low_lowering_zero} and
unity phase factors)
\begin{eqnarray*}
 |N_{-\alpha^a, -\alpha^a \rightarrow -2 \alpha^a} |^2 & = & -2 +
 \sum_{\Gamma,\Delta} (G^0)_{\Gamma\Delta}
 N_{-\alpha^a,0_{\Gamma}\rightarrow-\alpha^a}
 N^*_{-\alpha^a,0_{\Delta}\rightarrow-\alpha^a} \\
 & = & -2 + \sum_{\Gamma,\Delta} (G^0)_{\Gamma\Delta} (M^0)_{\Gamma
   a} (M^0)_{\Delta a} |N_{-\alpha^a, \alpha^a\rightarrow 0_a}|^2 \\
 & = & -2 + (M^0)_{aa} |N_{-\alpha^a, \alpha^a\rightarrow 0_a}|^2 =
 -2+2=0 
\end{eqnarray*}
while for $a\neq b$ it is simply
\begin{equation*}
 |N_{-\alpha^b, -\alpha^a \rightarrow - \alpha^a-\alpha^b} |^2 =
 \langle \alpha^b,-\alpha^a\rangle+0
\end{equation*}
since the negative simple root, $-\alpha^a$, can only be obtained from  
lowering zero weights by the $a$-th root. Further lowerings are
analogous to the ones in non-degenerate irreps. Combining all those
cases one finds a nice recursion relation for lowering
normalizations of the adjoint representation: 
\begin{equation}
 |N_{-\alpha^i, w_j\rightarrow (w_j-\alpha^i)}|^2 =
 \langle\alpha^i,w\rangle + |\langle w_j | w_i \rangle|^2
 \, |N_{-\alpha^i, (w_i+\alpha^i)\rightarrow w_i}|^2
\end{equation}
Again, for simplicity, phase factors are all taken to be unity.

Until now, we only specified lowering normalizations in non-degenerate
or adjoint representations. It is hard to find a general lowering
scheme but, fortunately, this is not needed as one can built up
arbitrary irreps from (multiple) tensor products of non-degenerate and
adjoint representations and transfer bases and normalization
constants. 

\subsubsection{Scalar products of basis states of adjoint zero weights}
\label{sec:scp}
We now return to the scalar product \eqref{expl_low_scp_zeros_adjoint} 
of the basis states of weight zero in the adjoint representation. From
\eqref{expl_low_def} and its Hermitian conjugate, we find 
\begin{eqnarray}
 \label{expl_low_scp_zeros}
 \langle 0_j| 0_i \rangle & = & \left (N_{-\alpha^i,\alpha^i\rightarrow
 0_i} N^*_{-\alpha^j,\alpha^j\rightarrow 0_j} \right)^{-1}
 \langle\alpha^j|E_{\alpha^j} E_{-\alpha^i} |\alpha^i\rangle \nonumber\\
 & = & \frac{1}{2} \, \langle\alpha^j| \left ([E_{\alpha^j},
   E_{-\alpha^i}] + E_{-\alpha^i} E_{\alpha^j} \right )
 |\alpha^i\rangle
\end{eqnarray}
where also \eqref{expl_low_n_simpl_r2} has been used. As
$\alpha^j-\alpha^i$ is an adjoint weight only if $i=j$, the commutator
$[E_{\alpha^j}, E_{-\alpha^i}]$ is non-vanishing only in the case
$i=j$ and is then given by
\begin{equation*}
 [E_{\alpha^i},E_{-\alpha^i}] = \langle\alpha^i,H\rangle.
\end{equation*}
For $i=j$, $E_{\alpha^i} |\alpha^i\rangle$ is zero because $2\alpha^i$
is never a root. Hence, in this case we find
\begin{equation}
 \label{expl_low_scp_zeros_ii}
 \langle 0_i| 0_i \rangle = \frac{1}{2}\, \langle\alpha^i|
 \langle\alpha^i,H\rangle |\alpha^i\rangle = \frac{2}{2}
 \langle\alpha^i|\alpha^i\rangle = 1
\end{equation}
On the other hand, for $i\neq j$, we are left with
\begin{equation}
 \langle 0_j| 0_i \rangle = \frac{1}{2}\, \langle\alpha^j|
 E_{-\alpha^i} E_{\alpha^j}|\alpha^i\rangle.
\end{equation}
We already know that in this case we have
\begin{equation*}
 E_{-\alpha^i} |-\alpha^j\rangle =
 \sqrt{\langle\alpha^i,-\alpha^j\rangle} \, |-\alpha^i-\alpha^j\rangle =
 \sqrt{|A_{ji}|} \, |-\alpha^i-\alpha^j\rangle
\end{equation*}
which can be used to fix the following raising normalizations:
\begin{equation}
 \label{expl_low_rais}
 E_{\alpha^i} |\alpha^j\rangle = \sqrt{|A_{ji}|} \,
 |\alpha^i+\alpha^j\rangle 
\end{equation}
It is now by means of \eqref{expl_low_rais} and its Hermitian
conjugate that we end up with
\begin{eqnarray}
 \label{expl_low_scp_zeros_ij}
 \langle 0_j| 0_i \rangle & = & \frac{1}{2}\, \langle\alpha^j|
 E_{-\alpha^i} E_{\alpha^j}|\alpha^i\rangle \nonumber\\
 & = & \frac{\sqrt{A_{ij}A_{ji}}}{2} \,
 \langle\alpha^i+\alpha^j|\alpha^i+\alpha^j\rangle = \frac{\sqrt{A_{ij}A_{ji}}}{2}
\end{eqnarray}
Eventually, the two cases in \eqref{expl_low_scp_zeros_ii} and
\eqref{expl_low_scp_zeros_ij} can be put together to what we wanted to 
prove, \eqref{expl_low_scp_zeros_adjoint}:
\begin{equation*}
 \langle 0_a|0_b\rangle = \frac{\sqrt{A_{ab}A_{ba}}}{2}
\end{equation*}
Note that this relation generalizes $\langle 0_a|0_b\rangle =
|A_{ab}|/2$ which is the one from Anderson \& Blazek \cite{Anderson:2001sd}. It
is apparent that their relation is restricted to the case where the
Cartan matrix is symmetric and, thus, applies only to Lie algebra
classes $A$, $D$ and $E$. In contrast, our relation applies to all Lie
algebras $A$ to $G$.

%%%%%%

\subsection{Decomposition algorithm}
\label{cgc_algorithm}
There are various algorithms for the computation of tensor product 
decompositions of simple Lie algebra representations. If it were only
for the Littlewood-Richardson coefficients there exists very efficient 
means. For instance, there is a fast algorithm based on characters and
Klimyk's formula \cite{Zol97}.\footnote{An implementation of this
 algorithm is found in a software called LiE which decomposes nearly
 any tensor product in less than a second \cite{Lie}.}   
However, not only are we interested in multiplicities but also in the
Clebsch-Gordan coefficients, that is, the coefficients of tensor
product states in terms of the states of the two irreps the tensor
product is taken of. We therefore proceed straightforwardly
by explicitly constructing states in the modules of the tensor product 
representation. This, basically, generalizes the Clebsch-Gordan
algorithm for $SU(2)$ which is well-known from the addition of angular
momenta. Such an explicit algorithm is also found in the works by Koh
et al.\ \cite{Koh:1984vs} and Anderson \& Blazek (\cite{Anderson:1999em},
\cite{Anderson:2000ni}, and \cite{Anderson:2001sd}) in which tensor
products of some lower-dimensional irreps of $E_6$ were computed. It
will become apparent that in this approach we will need knowledge of
consistent lowering normalizations and bases in the irreps the tensor
product is taken of. 

Before presenting the algorithm we still need another definition:
Dominant weights are weights in the product representation with
non-negative Dynkin coordinates only. They are of interest as they
can serve as highest weights of irreps in the
decomposition. (Apparently, highest weights of irreps are always
dominant weights.) Now, our Clebsch-Gordan algorithm is as follows:   
\begin{enumerate}
 \item Compute the highest weight as the sum of the highest weights of
   the two irreps. 
 \item Descend one level, order by weights and drop states until the
   remaining ones become linearly independent, i.e.\ they now form a
   basis. This is the CPU-intensive part. 
 \item Descend next level \ldots until the full irrep is
   constructed. 
 \item Find dominant weights and list them. 
 \item Compute basis of dominant weight states in the tensor product
   representation from the weights and their degeneracies in the two
   irreps the tensor product is taken of.
 \item Choose dominant weight with the most levels\footnote{For Lie
     algebras A to G there is an $r$-tuple, the so-called level
     vector, whose standard scalar product with the highest weight
     gives the level of the lowest weight, i.e.\ the maximal level
     \cite{Slansky:1981yr}.} and check if the complement (of dominant weights
   from step 4 with respect to the ones from step 5) is non-empty. If
   so, this is the highest weight of the next irrep in the
   decomposition. Otherwise, remove this dominant weight from the
   list and repeat step 6 as long as the list is non-empty. 
 \item Descend irrep whose highest weight is given by the dominant
   weight from step 6. Go to step 2\ldots 
 \item Algorithm ends automatically, check dimensions.
\end{enumerate}

As an illustration of the algorithm, let us consider the following
neat example for $SU(3)$:
$$\bf 3\otimes\bar{3} = ?$$
Both the $\bf 3$ and $\bf\bar{3}$ are non-degenerate and, thus,
lowering normalizations are taken from \eqref{expl_low_non_deg_irrep} 
\begin{eqnarray*}
 \bf 3: & \qquad & |10\rangle \stackrel{1}{\rightarrow} 1\cdot
 |\bar{1}1\rangle \stackrel{2}{\rightarrow} 1\cdot |0\bar{1}\rangle
 \\ 
 \bf \bar{3}: & \qquad & |01\rangle \stackrel{2}{\rightarrow} 1\cdot
 |1\bar{1}\rangle \stackrel{1}{\rightarrow} 1\cdot |\bar{1}0\rangle.
\end{eqnarray*}
Here, numbers over arrows denote the lowering operators corresponding
to the respective simple roots. The highest weight in the tensor
product representation is given by  
$$|11\rangle = |10\rangle |01\rangle$$
from which we descend, using \eqref{tp_cartan_low_rais}, and build up 
the essential part of what turns out to be an octet: 
\begin{eqnarray*}
 |11\rangle & \stackrel{1}{\rightarrow} & |\bar{1}2\rangle =
 |\bar{1}1\rangle |01\rangle \stackrel{2}{\rightarrow} \sqrt{2}
 \cdot |00_2\rangle = |0\bar{1}\rangle |01\rangle +
 |\bar{1}1\rangle |1\bar{1}\rangle \stackrel{1,2}{\rightarrow} \ldots \\
 |11\rangle & \stackrel{2}{\rightarrow} & |2\bar{1}\rangle =
 |10\rangle |1\bar{1}\rangle \stackrel{1}{\rightarrow} \sqrt{2}
 \cdot |00_1\rangle = |10\rangle |\bar{1}0\rangle +
 |\bar{1}1\rangle |1\bar{1}\rangle \stackrel{1,2}{\rightarrow}
 \ldots 
\end{eqnarray*}
We need not descend further as dominant weights cannot occur below the
level with the zero weight. Dominant weight states other than
$|11\rangle$ are  
\begin{eqnarray*}
 \sqrt{2} \cdot |00_1\rangle = |10\rangle |\bar{1}0\rangle +
 |\bar{1}1\rangle |1\bar{1}\rangle \\ 
 \sqrt{2} \cdot |00_2\rangle = |0\bar{1}\rangle |01\rangle +
 |\bar{1}1\rangle |1\bar{1}\rangle.
\end{eqnarray*}
Now we list basis vectors for the $|00\rangle$-space  --- $|10\rangle
|\bar{1}0\rangle, |\bar{1}1\rangle |1\bar{1}\rangle, |0\bar{1}\rangle
|01\rangle$ --- which turns out to be of dimension 3. Having already
two states of this weight there can only be one additional state which
is orthogonal to the first two. We find
$$\sqrt{3} \cdot |00\rangle = |10\rangle |\bar{1}0\rangle -
|\bar{1}1\rangle |1\bar{1}\rangle + |0\bar{1}\rangle |01\rangle$$  
which, in fact, turns out to be a singlet. Note that all kets are
normalized to unity. The result is well known,
$$\bf 3\otimes\bar{3} = 8+1.$$

It is by the very nature of this algorithm that for larger Lie
algebras (rank $>2$) or representations with higher dimensions ($>10$)
computations grow rapidly in length and therefore become very
tedious. The implementation of this algorithm is described in the
following sections. In principle --- that is, up to restrictions
coming from memory and CPU-time demands --- it enables us to compute 
any tensor product decomposition of irreps of Lie algebras $A$ to $G$,
as well as the corresponding Clebsch-Gordan coefficients. Note,
however, that in general Clebsch-Gordan coefficients are not universal
when arbitrary irreps with degenerate weight spaces are involved as
they only make sense if, at the same time, bases in the degenerate
weight spaces are specified. 

%%%%%

\newpage
\section{Overview of the software structure}
\label{s:overview}
Our code is divided into the following parts:
\begin{center}
   \begin{tabular}{ l |  p{9cm}}
   File & Summary \\
   \hline
   \verb=Aux.ml= & Auxiliary functions such as list manipulation
   routines (beyond the ones from \texttt{O'Caml} standard module
   \verb=List=), combinatorical functions needed in
   \verb=Liealg.ml= or simplification methods for \verb=big_int=
   numbers\ldots \\  
   \verb=Liealg.ml= & Subprogram that is capable of computing weight
   lists and weight degeneracies \\
   \verb=Alg.ml= & Subprogram to model an algebra \\
   \verb=CGC.ml= & Main program for decomposing tensor product
   representations and computing Clebsch-Gordan coefficients \\ 
   \verb=CGC_nb.ml= & User notebook specifying input and output of
   the computation.
   \end{tabular}
\end{center}
Our software makes use of the modules \verb=List= and \verb=Array=
from the \texttt{O'Caml} standard library. Moreover, module \verb=Num=
from the \verb=Num= library is used for arbitrary-precision integer /
rational arithmetic. 

It is by the very nature of representation theory that the complexity
of weight spaces in tensor product representations grow rapidly when
larger irreps or irreps of larger algebras are involved in the tensor
product representation. In such cases, it often does not make sense to 
output the resulting Clebsch-Gordan coefficients. Instead, one better
proceeds until computations reach a much simpler result which may then
be a suitable output. This is the reason why our program cannot be
distributed as a single compiled user-friendly binary executable. It
should rather be understood as a package that provides the tools to do 
computations that require knowledge of Clebsch-Gordan
coefficients. We therefore included a notebook file \verb=CGC_nb.ml=
where such input and output can be specified.

\section{Description of the individual software components}
\label{s:components}
\subsection{Aux}
The module \texttt{Aux.ml} contains auxiliary functions and
operations needed for the explicit construction of irreps and
the Clebsch-Gordon decomposition of multiple tensor products. Parts
of it bear similarities to the corresponding modules on lists in the
\texttt{O'Mega} matrix element generator for quantum-field theoretical 
amplitudes~\cite{omega}. This module contains e.g.\ functions to
generate a list of all inequivalent pairs of numbers smaller than
$n$. Most functions in \verb=Aux= are rather primitive and may go
without further explanation.

We should, however, mention that here we define the following tree
structure,\\  
\verb~type 'a tree = Empty | Node of int * 'a * 'a tree * 'a tree~,\\
which will be used in the construction of multiple tensor products
(see\\ \verb~class irrep_in_tp~ in \verb~CGC.ml~). Some tree
manipulations are also contained in this module.

%%%%%

\subsection{Liealg}
\label{s:c:liealg}

This subprogram generates by explicit construction the list of all
possible weights of any irrep in any simple Lie algebra. It determines
all the states of an irrep either in the Dynkin basis or as a tuple of
numbers how many times one has to descend from the (unique) highest
weight to get to the corresponding weight of the irrep. A trivial
piece of information is the level, which is just the number of total
descends from the highest weight. The construction proceeds simply by
consecutive subtraction of the rows of the Cartan matrix from the
highest weight. In that construction, some states are constructed more
than once (if they can be reached on different paths of descent from
the highest weight); multiple states are eliminated and the states at
each level will be sorted.

The recursive Freudenthal formula
(cf. (\ref{eq:freudenthalabstract}),(\ref{eq:freudenthal})) is used to
determine the degree of degeneracy of each level, an information of
importance for the explicit generation of the space of states of each
irrep.  

Here, we briefly describe the setup of the elementary objects and most
important functions of our software implementation in the module
\texttt{Liealg}. A weight is denoted by a pair, where the first entry
itself is a pair, giving the level in an irrep starting with zero for
the highest weight, the second entry is a list of integers which has as
$n$th entry the number of how many times the simple root $\alpha^n$
has been applied to the highest weight. The second entry of the weight
pair is again an integer list, namely the Dynkin coefficients.  
The function
\verb=complete_descent: liealgebra -> =\\
\verb=            int list -> ((int * int list) * int list) list=\\
has as arguments a Lie algebra (being one of the elements 
\texttt{A of int}, \texttt{B of int}, \texttt{C of int}, \texttt{D of
  int},  \texttt{E6}, \texttt{E7}, \texttt{E8}, \texttt{F4}, or
\texttt{G2}) and an integer list giving the Dynkin coefficients of the
highest weight of the representation. The result is the list of all
weights of the irrep. 

This set of all weights is used by the function \\
\verb=freudenthal: liealgebra -> int list -> (int * (int * int list) =\\
\verb=     * (((int list * int) * int) * (int * int list)) list) list=\\
which has as input again the Lie algebra and a highest weight of an
irrep, and gives back a list of all the weights (degeneracy, level
and Dynkin coefficient), and a list of all states relevant for the
recursive calculation according to the sum on the right-hand side of
(\ref{eq:freudenthal}). It is a constructive recursive function
implementing the Freudenthal algorithm. 

These two functionalities are combined in the function: \\
\verb=full_descent: liealgebra -> int list -> =\\
\verb=            ((int * int * int list) * int list * int) list=\\
which gives back just the list of weights for the irrep defined by the
highest weight of type \verb=int list= including the
degeneracies. Each weight is augmented by the Dynkin number that
corresponds to the zeroth root. 

When using the \texttt{O'Caml} toplevel a more readable output of the weight
system is given by means of the function\\
\verb~weights: Liealg.liealgebra -> int list -> unit~\\
where again the two input parameters correspond to the Lie algebra and
the heighest weight.

We just mention the function\\
\verb=cartan: liealgebra -> int list list=\\
which gives the Cartan matrix for the corresponding Lie algebra. 

%%%%%

\subsection{Alg}

The construction of explicit states in the tensor product
representation requires some linear algebra techniques most of which
are based on a suitable implementation of the Gaussian
algorithm. However, as we aim at producing exact Clebsch-Gordan
coefficients --- as opposed to float numbers --- we first have to
model objects that represent numbers that are linear combinations of
square roots of non-negative integers with rational coefficients,
i.e.\  
\begin{equation}
\label{s.c.lcrs}
  \sum_{n=0}^{\infty} q_n \sqrt{n}
\end{equation}
where all but finitely many $q_n\in\mathbb{Q}$ are zero. We then
extend the numbers in \eqref{s.c.lcrs} to a field and also model the
notion of a vector space over this field. Our linear algebra routines
are then based on the latter two.

We have the following classes: 
\begin{itemize}
 \item \verb=class rational=:\\
   Rational numbers $\mathbb{Q}$ based on
   the arbitrary-precision rational numbers from the \verb=Num=
   library. Its constructor takes two arguments referring to the two 
   signatures \verb=(numerator:int)= and
   \verb=(denominator:int)=. Commonly used 
   methods are:\\
   \verb=add: rational -> rational=\\
   \verb=multiply: rational -> rational=\\
   \verb=divide_by: rational -> rational=\\
   \verb=is_zero: -> bool=\\
   \verb=is_bigger_than_zero: -> bool=\\
   \verb=print: -> string=\\

 \item \verb=class linear_comb_rational_sqrts=:\\
   Models algebra over the
   rational numbers generated by the square roots of non-negative
   integers, see \eqref{s.c.lcrs}. Its constructor takes an argument
   of type \verb=(rational * int) list= where every list entry
   corresponds to a summand in \eqref{s.c.lcrs}. Important methods
   are:\\ 
   \verb=get_list: -> (rational * int) list=\\
   \verb=add: linear_comb_rational_sqrts -> linear_comb_rational_sqrts=\\
   \verb=multiply: linear_comb_rational_sqrts -> =\\
   \verb=                 linear_comb_rational_sqrts=\\
   \verb=remove_zeros: -> linear_comb_rational_sqrts=: To remove list
   entries that correspond to zeros in \eqref{s.c.lcrs} which can be 
   remnants of methods \verb=add= and \verb=multiply=.\\
   \verb=symplify: -> linear_comb_rational_sqrts=: Simplifying by
   extracting roots as much as possible and simplifying terms that
   include identical square roots.\\
   \verb=is_zero: -> bool=\\
   \verb=print: -> string=: String output to be used in the \texttt{O'Caml} toplevel.\\
   \verb=tex: -> string=: Outputs number in \TeX  format.\\
   \verb=export_math: -> string=: Outputs number in Mathematica format.\\

 \item \verb=class lcrs_field=:\\
   Models the field extension of 
   \verb=linear_comb_rational_sqrts= numbers via the inclusion of
   fractions of \verb=linear_comb_rational_sqrts= numbers. Its
   constructor takes two \verb=linear_comb_rational_sqrts= objects
   referring to the numerator and denominator. Methods of importance
   are:\\
   \verb=get_num: -> linear_comb_rational_sqrts=: Yields the numerator
   of the fraction.\\
   \verb=get_denom: -> linear_comb_rational_sqrts=: Yields the
   denominator of the fraction.\\
   \verb=remove_zeros: -> lcrs_field=: Removes zeros in both the
   numerator and the denominator.\\
   \verb=symplify: -> lcrs_field=: Symplifies the numerator and the
   denominator. In the case where the denominator consists only of one
   term further simplifies the fraction.\\
   \verb=add: lcrs_field -> lcrs_field=\\
   \verb=multiply: lcrs_field -> lcrs_field=\\
   \verb=invert: -> lcrs_field=: If non-zero, gives the inverse.\\
   \verb=is_zero: -> bool=\\
   \verb=is_bigger_than_zero: -> bool=\\
   \verb=print: -> string=\\
   \verb=tex: -> string=\\
   \verb=export_math: -> string=\\

 \item \verb=class ['a] lcrs_vector=:\\
   Models the vector space over
   \verb=lcrs_field= with basis vectors labeled by values of type
   \verb='a=. The constructor takes an argument of the following type: 
   \verb=(lcrs_field * 'a) list= where the list represents a linear
   combination of basis vectors of type \verb='a= with
   \verb=lcrs_field= coefficients. Useful methods of this class
   are:\\
   \verb=add: lcrs_field -> 'a -> 'a lcrs_vector=: Adds another term
   to the list.\\
   \verb=add2: 'a lcrs_vector -> 'a lcrs_vector=: Adds another vector
   of the same type \verb='a= and returns the sum.\\
   \verb=get_list: -> (lcrs_field * 'a) list=\\
   \verb=print: -> (string * 'a) list=\\
   \verb=scale: -> lcrs_field -> 'a lcrs_vector=: Scales the vector by
   a number of our field.\\
   \verb=remove_zeros: -> 'a lcrs_vector=\\
   \verb=symplify: -> 'a lcrs_vector=\\
   \verb=is_zero: -> bool=\\
\end{itemize}

Based on \verb=lcrs_field=, we now have the following functions:
\begin{itemize}
 \item \verb=number: int -> int -> int -> lcrs_field=:\\
   Instantiates
   (via \texttt{number a b n}) objects of \verb=lcrs_field= that represent
   numbers of the simple form $\frac{a}{b} \sqrt{n}$.
 \item \verb=gcd_of_lcrs_fields: lcrs_field list -> lcrs_field=:\\
   Computes the greatest common divisor of a set of \verb=lcrs_field=
   numbers.
 \item \verb=array_product: .. array -> .. array -> lcrs_field=:\\ This
   function takes two arguments that e.g.\ are of type\footnote{This
     function is polymorphic in that the arguments are only restricted
     to be an array of an object that comes with suitable methods
     \verb~add~ and \verb~multiply~. This is what we mean by \verb~..~
     above.} \verb=lcrs_field array= and returns the scalar product
   provided that the two arrays have equal length.  
 \item \verb=matrix_multiply: .. array array -> .. array array ->=\\
   \verb=             .. array array=: \\
   Multiplies e.g.\ two matrices of type
   \verb=lcrs_field array array= and yields a matrix of the same type,
   provided that dimensions of matrices fit.    \\
 \item \verb=gauss: lcrs_field array array -> lcrs_field array array=\\
    \verb=-> lcrs_field array array=:\\ Gaussian algorithm (adapted
    from \cite{carette}) transforming a \verb=lcrs_field= matrix to
    Gaussian shape. The second argument is meant to be a column vector
    representing the right-hand side of a system of linear equations.
 \item \verb=solve: ... array array -> lcrs_field array=:\\ Gives an
   explicit arbitrary solution (provided there is one) to a system
   of linear equations which has already been brought into Gaussian
   shape. Here, the argument \verb=...= (which represents the system
   of linear equations) is again to some extent polymorphic. It is 
   sufficient to state that e.g.\ the output of \verb=gauss= is of the
   right type.  
 \item \verb=invert: lcrs_field array array -> lcrs_field array array=:\\
   Inverts (if possible) a square matrix over \verb=lcrs_field=. Note
   that although \texttt{invert} is not needed in computing Clebsch-Gordan
   coefficients, it is useful when it comes to basis changes in
   \verb=CGC_nb=.  
 \item \verb=linearly_dependent: .... list -> bool=:\\ Tests whether or
   not a set / list of vectors are linearly dependent. The argument \verb=....=
   is again polymorphic. As an example, an argument could be of type
   \verb=['a] lcrs_vector list=. Note that
   although this function is not needed in computing Clebsch-Gordan 
   coefficients it proves useful in constructing bases etc.
\end{itemize}

%%%%%

\subsection{CGC}

This part contains the algorithm to explicitly decompose tensor
products of irreducible representations of Lie algebras $A$ - $G$. 

We have the following functions / variables:
\begin{itemize}
 \item \verb~level_vector: Liealg.liealgebra -> int list~:\\
   Returns the level vector of the Lie algebra. They are taken
   from \cite{Slansky:1981yr}. As already stated in
   Section \ref{cgc_algorithm} dominant weights in the tensor product
   representation are sorted according to their maximal levels, the
   latter of which can be computed by means of the level vector.   
 \item \verb~adjoint: Liealg.liealgebra -> int list~:\\
   Returns the highest weight of the adjoint representation for a
   given Lie algebra. Generically, we only have consistent lowering
   normalizations for non-degenerate or adjoint irreps, see
   Section \ref{explicit_lowering}. It is by means of this function,
   \texttt{adjoint}, that we can recognize adjoint
   representations.\footnote{Note that in 
     general degeneracies are not sufficient to single out adjoint
     representations.}
 \item \verb~empty_input = ([], (([], []), []))~:\\ 
   When instantiating an object of the class \texttt{irrep} (see
   below) the variable \texttt{empty\_input} as an input parameter
   indicates that the standard lowering normalizations are
   used. This, of course, works in the case of non-degenerate or
   adjoint representations only. 
 \item \verb~scp_zero_weights: Liealg.liealgebra -> int -> int~\\
   \verb~-> Alg.lcrs_field~:\\
   Gives the scalar products of the basis states of the zero weight
   subspace in the adjoint representation from the Cartan matrices
   according to \eqref{expl_low_scp_zeros_adjoint}. Our basis states
   are defined in terms of descending the simple root states. The
   second and third argument of type \texttt{int} are restricted from
   $1$ to the rank of the Lie algebra. The knowledge of those scalar
   products are needed for consistent lowering normalizations for
   adjoint representations, see \eqref{expl_low_lowering_zero}.
\end{itemize}

We have the following classes:
\begin{itemize}
 \item \verb~class ket~:\\
   Models a generic state. Its constructor takes two arguments of
   which the first is an \verb=int list= to specify the weight of the
   state and the second is of type \verb=int= to label
   degeneracies. One of its methods are:\\ 
   \verb=print: -> string=: Outputs a string containing the weights
   and the degeneracy label.\\
 \item \verb~class irrep~:\\
   This class is the data structure of irreps that is needed to
   compute their tensor products. Its constructor takes as arguments the
   Lie algebra of type \verb=Liealg.liealgebra=, the highest
   weight of type \verb=int list=, and a third input parameter of type
   \verb=b * ((c * d) * e)= where\\
   \verb~b = '_a list~,\\
   \verb~c = (int * ket) list~,\\
   \verb~d = ((int * int) * int Alg.lcrs_vector) list~,\\
   \verb~e = ((int * int) * Alg.lcrs_field) list~.\\
   In the case of non-degenerate or adjoint irreps consistent lowering
   normalizations and scalar products can be computed as discussed in
   Section \ref{explicit_lowering}. The weight systems are obtained from the
   subprogram \texttt{Liealg}, cf. Section \ref{s:c:liealg}. In these
   cases the third input parameter can be chosen to be
   \texttt{empty\_input}. On the other hand, more 
   complicated irreps first need to be constructed in a simpler
   tensor product. Subsequently, one reads off consistent lowering 
   normalizations and scalar products which finally are to be used as
   an input parameter upon initializing an object of type
   \texttt{irrep}. In this case, the second argument specifying the
   highest weight is redundant and is ignored.\footnote{It suffices
     to put in an empty list \verb~[]~.} Here, type \verb=b= becomes 
   \verb=(int * lcrs_ketket) list=,\, i.e.\ this part is the list of
   pairs of an 
   integer enumerating all weights and the corresponding states in the
   tensor product representation of type \verb=lcrs_ketket= (to be defined
   below). Type \verb=c= is meant to model the list of pairs of the
   enumerating index and the ket (weights plus degeneracy
   index). Next, data of type \verb=d= encodes the lowering
   normalizations: For every list entry, the second integer gives the 
   simple root by which we lower the state specified by the
   first integer; the object of type \verb=int Alg.lcrs_vector=
   gives the resulting linear combinations in terms of the unique
   state labels. Finally, type \verb=e= encodes the scalar products:
   For every list entry, the two integers specify the states of which
   the scalar product is taken; the result is of type
   \verb=Alg.lcrs_field=.\\

   \noindent Its methods are:\\
   \verb=get_la: -> Liealg.liealgebra=: Returns the Lie algebra the
   irrep belongs to.\\
   \verb=get_hw: -> int list=: Returns the highest weight of the irrep.\\
   \verb=get_dim: -> int=: Returns the dimension of the irrep.\\
   \verb=is_adjoint: -> bool=: Yields true if the irrep is adjoint and
   false otherwise.\\
   \verb=get_list: -> (int * ket) list=: Returns the list that
   contains the unique labeling of the kets.\\
   \verb=print_list: -> (int * string) list=: Like \verb=get_list= but
   expresses kets as \verb=string=.\\
   \verb=get_ket: int -> ket=: Returns the ket specified by the
   integer.\\
   \verb=get_scalar_product: int -> int -> lcrs_field=:\\ 
   Returns the
   scalar products of the states specified by the two integers. If the
   object of \texttt{irrep} has been initialized by means of a
   non-trivial third argument resulting from another tensor product
   the scalar products are read off from this parameter. Otherwise,
   scalar products are computed as discussed in Section
   \ref{explicit_lowering}.\\ 
   \verb=lower: int -> int -> int lcrs_vector=: Returns the linear
   combination that results from lowering the state specified by the
   second integer with the simple root denoted by the first
   integer. Again, lowering normalizations are read off from a
   non-trivial third input parameter of the constructor or follow from
   the rules for non-degenerate or adjoint representations discussed
   in Section \ref{explicit_lowering} if the third input parameter is
   trivial.\\ 

 \item \verb=class ['a] lcrs_2vectors=:\\
   This class is an inheritor of the class \verb=['a] lcrs_vector=
   where \verb='a= is taken to be \verb='a * 'a=. Objects of this
   class can be viewed as elements of the tensor product of two
   \verb='a= vector spaces.\\

 \item \verb=class lcrs_ketket=:\\
   This class models states in the tensor product representation of
   two irreps. It is inherited from the class 
   \verb=[int] lcrs_2vectors=. Its constructor takes one argument of
   type \verb=(Alg.lcrs_field * (int * int)) list = which is meant to
   be a linear combination over \verb=lcrs_field= of basis states
   labeled by two integers which specify the respective
   states in the two irreps the tensor product is taken of.\\

   \noindent We have the following methods (in addition to the
   inherited ones):\\ 
   \verb=add_lcrs_ketket: lcrs_ketket -> lcrs_ketket=:\\
   \verb=weight: irrep -> irrep -> int list=: Specifying the two
   irreps the tensor product is taken of, this method yields the
   weight of the state. Note that in order to save RAM an object of
   \texttt{lcrs\_ketket} does not know which irrep it belongs to which
   is why such information must be given here (and in what follows) as
   an input parameter.\\
   \verb=print_kets: irrep -> irrep -> (string * (string * string)) list=:\\ 
   \verb=print_to_string: irrep -> irrep -> string=:\\
   \verb=lower: b -> b -> int -> lcrs_ketket=: where \verb=b=
   abbreviates\\
   \verb~b = (int -> int -> int Alg.lcrs_vector)~ which is precisely
   the type of method \verb=lower= of class \verb=irrep=. In fact, the 
   first two arguments should be supplied by the data from method 
   \verb=lower= of \verb=irrep= for the two irreps of the tensor
   product. The third input parameter specifies the simple root that 
   acts as lowering operator.\\
   \verb=scalar_product: lcrs_ketket -> b -> b -> Alg.lcrs_field=:
   where \verb=b= abbreviates \verb~b = (int -> int -> Alg.lcrs_field)~ 
   which is the type of method
   \verb=get_scalar_product= of class \verb=irrep=. Also here, the
   second and the third argument should be supplied by the method
   \verb=get_scalar_product= of the two irreps of the tensor
   product. The first argument is another object of
   \texttt{lcrs\_ketket} with which the scalar product is to be
   computed.\\ 

 \item \verb=class irrep_in_product_rep:=\\
   This class models irreps that appear in the tensor product. Its
   constructor takes only one argument of type \verb=lcrs_ketket=
   which corresponds to the highest weight of the irrep. \\

   \noindent We have the following methods:\\
   \verb=get_hw: -> lcrs_ketket=: Returns the highest weight.\\
   \verb=get_list: -> lcrs_ketket list list=: Returns the product
   states in the irrep grouped according to their level.\\
   \verb=get_dim: -> int=: Returns the dimension of the irrep.\\
   \verb=print: irrep -> irrep -> =\\
   \verb=              (string * (string * string)) list list list=:\\ 
   \verb=print_flattened: irrep -> irrep -> =\\
   \verb=              (string * (string * string)) list list=:\\
   \verb=dominant_weights: irrep -> irrep -> lcrs_ketket list=:
   Returns the states whose weights are dominant.\\
   \verb=descend: irrep -> irrep -> irrep_in_product_rep=: After
   instantiating an object of class \texttt{irrep\_in\_product\_rep} 
   the full irrep is constructed by means of the method
   \texttt{descend}. Again the two irreps of the tensor product must
   be supplied as input parameters.\\
   \verb=prepare: irrep -> irrep -> b * ((c * d) * e)=: Reads out
   consistent lowering normalizations and scalar products for an
   irrep in the tensor product and prepares this data as an input
   parameter for the constructor of class \texttt{irrep}. It is by
   this mechanism that one can in principle instantiate objects of
   class \texttt{irrep} that model arbitrary irreps. Here, type
   abbreviations read: \\
   \verb~b = (int * lcrs_ketket) list~,\\
   \verb~c = (int * ket) list~,\\
   \verb~d = ((int * int) * int Alg.lcrs_vector) list~,\\
   \verb~e = ((int * int) * Alg.lcrs_field) list~.\\
   In words, variables of these types contain the association of labels
   to states in the tensor product representation\footnote{This bit of
   information is not needed in the constructor for class \verb~irrep~
   which is why there this parameter is polymorphic.}, the association
   of labels to the weights, consistent lowering normalizations, as
   well as the scalar products of the chosen basis states,
   respectively.\\

 \item \verb=class clebsch_gordan_decomposition:=\\
   This class is the data structure for the Clebsch-Gordan
   decomposition. It contains as its main part an implementation of
   the decomposition algorithm from Section \ref{cgc_algorithm}. The
   constructor takes two arguments of type \texttt{irrep} that specify
   the two factors of the tensor product. Upon initialization only the
   highest weight representation in the product representation is
   computed.\\

   \noindent Its methods are:\\
   \verb=get_irrepa: -> irrep=: Returns the first factor in the tensor
   product.\\
   \verb=get_irrepb: -> irrep=: Returns the second factor in the
   tensor product.\\
   \verb=get_irreps_in_product: -> irrep_in_product_rep list=: Lists
   all irreps that appear in the tensor product.\\
   \verb=basis_product: int list -> lcrs_ketket list=: Computes a
   (non-ortho\-gonal) basis for the tensor space for a given weight.\\
   \verb=decompose: -> unit=: Completely decomposes the tensor product
   by constructing states explicitly.\\
   \verb=check_dims: -> bool=: Yields true if dimensions of the tensor
   product and its decomposition match, otherwise returns false.\\
   \verb=result: -> string=: Returns a string output of the
   decomposition.\\

 \item \verb~class irrep_in_tp~:\\
   This class has been designed in view of multiple tensor
   products: Its objects (corresponding to irreps) can be tensorized
   to give a certain irrep in the tensor product which becomes again
   an object of \verb=irrep_in_tp=. Hence, an iteration of this
   process is possible. We use a certain tree structure (\texttt{int
   tree}) the internal variable of this class in order to circumvent
   the type definiteness of classes in \texttt{O'Caml}. The
   constructor of this class takes an argument of type \texttt{irrep}.\\

   \noindent Its methods are:\\
   \verb=get_irrep: -> irrep=: Returns the irrep the object belongs to.\\
   \verb=get_res: -> (int * int tree Alg.lcrs_vector) list=:\\
   Expresses states in this irrep in terms of the states of the
   multiple tensor product: It returns a list of pairs where the first
   entry labels the state in the irrep while the second entry in the
   pair is a linear combination of the tree that encodes the basis
   states of the multiple tensor product.\\
   \verb=otimes: irrep_in_tp -> int -> irrep_in_tp=: Tensorizes the
   irrep with another irrep and returns another object of type
   \verb=irrep_in_tp= that belongs to the irrep specified by the
   second argument according to the order in which irreps in the
   tensor product are constructed (\verb=1= gives the highest weight
   irrep, \verb=2= the next one\ldots).\\
   \verb=expand: int -> int tree Alg.lcrs_vector=: Returns the state
   specified by the first argument in terms of a linear combination of
   trees.\\
   \verb=untree: (int * string Alg.lcrs_vector) list=: Lists all
   states in the irrep as a pair of the state label and a linear
   combination of bases in the tensor product which are represented as
   strings.\\  
   \verb=filter: int -> int list -> irrep_in_tp=: Filters out only
   terms that include product bases where numbers from \texttt{int list}
   appear in the place that corresponds to the tensor product factor
   specified by the first argument.\\ 
   \verb=chbasis: int -> (int * int Alg.lcrs_vector) list ->=\\
   \verb=                                   irrep_in_tp=:\\ 
   Performs a basis change in the tensor product factor specified by
   the first argument. The linear transformation is given by the
   second argument by means of associating to every state a linear 
   combination of states.\\ 
   \verb=is_sym: int -> int -> int=: Tests the symmetry under exchange
   of factors in the tensor product specified by the first two
   arguments. Yields \verb=1,-1,0= in the case of symmetry,
   antisymmetry, or indefinite symmetry, respectively.\\
   \verb=scale: Alg.lcrs_field -> irrep_in_tp=: Scales all states by
   the first argument.\\
   \verb=tensor: int -> int list -> Alg.lcrs_field=: Gives for the
   state specified by the first argument, the coefficient that belongs
   to a product basis state specified by the second argument.
\end{itemize}

Furthermore, we included some functions that can be useful in discussing
symmetry breaking in particle physics by means of a generalized Higgs
mechanism: Invariant higher-dimensional operators (i.e.\ singlets in
the multiple tensor products) may contain what we call Higgs fields
which means that they are supposed to acquire a vacuum expectation
value (vev) in a certain direction in the weight spaces that
correspond to the irreps the Higgs transform under. Upon vev insertion
higher-dimensional operators yield an effective potential which one
may want to compute. The stability groups of those vevs determine the
subsymmetry to which the full symmetry is said to be spontaneously
broken. It then makes sense to express the effective potential in
terms of irreps of the subsymmetry. Here, we do not aim at giving a full
set of routines that are capable of doing such computations in
general. We rather give some general functions that we found useful
and present one additional function that was well suited to our
problem though its extension to similar problems should be possible. 
\begin{itemize}
  \item \verb~e_lower: irrep -> int -> int Alg.lcrs_vector~\\
    \verb~-> int Alg.lcrs_vector~: This function models the lowering
    operator that is associated to the simple root specified by the
    second argument of type \texttt{int}. Given an irrep and a simple
    root it can be viewed as an operator that acts on an object of
    type \verb=int Alg.lcrs_vector= and returns another object of this
    kind. 
  \item \verb~comm: (int Alg.lcrs_vector -> int Alg.lcrs_vector)~\\
    \verb~-> (int Alg.lcrs_vector -> int Alg.lcrs_vector)~\\
    \verb~-> int Alg.lcrs_vector -> int Alg.lcrs_vector~:\\ This
    function gives the commutator of two lowering operators. Such a
    function is useful because lowering operators that correspond to
    arbitrary roots are in general given by such multiple
    commutators. For instance, one may want to break to a maximal
    subalgebra that is built from the zeroth root. A vev state should
    then be invariant under the multiple commutator of lowering
    operators that corresponds to the zeroth root.\\
  \item \verb~scp: irrep -> int Alg.lcrs_vector -> int Alg.lcrs_vector~\\
    \verb~-> Alg.lcrs_field~: Computes the scalar product of two
    states that belong to a given 
    irrep.\\
  \item \verb~scalar_products: irrep -> int -> int~\\
    \verb~-> int Alg.lcrs_field array array~:
    Prints scalar products of weight states in irrep. The first
    integer argument specifies weight in irrep while the second
    integer specifies how many further weights are to be selected. Use
    \verb~Aux.print_array~ to nicely display result.\\
  \item \verb~gram: irrep -> int Alg.lcrs_vector list~\\
    \verb~-> int Alg.lcrs_vector list -> int Alg.lcrs_vector list~:\\
    Performs a partial Gram-Schmidt procedure. Given a subset of
    orthogonal states in a basis (second argument), it subtracts for
    the remaining states in the basis (third argument) their
    projections on the orthogonal states. Then, the original set of
    orthogonal states and the complement are orthogonal.\\
  \item \verb~chbasis_list: int Alg.lcrs_vector list -> int~\\
    \verb~-> (int * int Alg.lcrs_vector) list~:
    The function \verb~chbasis_list~ returns the basis transformation
    rule in the format that is needed in the second argument of 
    \texttt{class\ irrep\_in\_tp}, method \texttt{chbasis}. It
    computes how the generic weight states in the irrep, specified by
    an offset parameter of type \texttt{int} (second argument) and the
    length of the basis (first argument), are expressed in terms of
    the new basis vectors. 
  \item \verb~eff_couplings: irrep_in_tp -> irrep_in_tp~\\
    \verb~-> (int list * string) array -> unit~:\\
    This function is specialized to the case where one wants to
    compute the effective potential that results from an operator /
    multiple tensor product that consists of three identical irreps
    and another irrep that is said to acquire a vev which is left
    invariant by some subsymmetry.\footnote{We were interested in the
      case of the tensor product $\bf 27\otimes 27 \otimes 27 \otimes
      650$ of $E_6$ that upon vev insertion yields an effective
      renormalizable superpotential.} The output is the effective
    superpotential in 
    terms of irreps of the subalgebra. The first argument is a singlet 
    in the fourfold tensor product while the second one takes the
    singlet of the threefold tensor product (the fourfold product
    without the Higgs irrep). The latter is used as a means to
    recognize the terms that are invariant under the subsymmetry. The
    third argument is an array that contains the branching of the
    irrep with respect to the subsymmetry in that each array entry
    associates a label of type \verb~string~ to a subset specified 
    by the integer labels of weight states. The output consists of a
    linear combination of terms that are separately invariant under
    the subsymmetry. In this expression, every term is an abbreviation
    of the corresponding terms in the threefold tensor product (second 
    argument).
\end{itemize}

\subsection{CGC\_nb}
This part should be understood as a notebook file that is supplied
with the computations one is interested in. Here, we include the
three examples for testing purpose presented in Section \ref{s:test}.

\newpage
\section{Installation instructions}
\label{s:installation}
\subsection{Toplevel mode}
Start \texttt{O'Caml} toplevel via:
\begin{verbatim}
ocaml nums.cma Aux.cmo Liealg.cmo Alg.cmo CGC.cmo
\end{verbatim}
It is convenient to write operations in a notebook file
``\verb+CGC_nb.ml+'' which can also be loaded into the
toplevel\footnote{Of course, this can also be done in the step before.}:
\begin{verbatim}
#use "CGC_nb.ml";;
\end{verbatim}

Functions / objects in modules ``\verb+Aux+'', ``\verb+Liealg+'',
``\verb+Alg+'', ``\verb+CGC+'', and ``\verb+CGC_nb+'' can be accessed
as follows: 
\begin{verbatim}
open Aux;;
...
\end{verbatim}

\subsection{Compilation mode}
Specify notebook file (e.g.\ \verb+CGC_nb.ml+) and output executable in
\verb+Makefile+ and run make.
\begin{verbatim}
 make
\end{verbatim} 
Then, run executable.

Note that the generation of weight systems (\verb=Liealg.ml=) can be
regarded as a subprogram of its own. As here input parameters are
limited to the specification of the Lie algebra and the heighest
weight of the irrep, a standalone executable can be useful. We
therefore included an option in the \verb=Makefile= that compiles
\verb=Liealg.ml= and \verb=Liealg_ex.ml= the latter of which specifies
the input interface. The \verb=make= command reads
\begin{verbatim}
  make lie.{opt|bin}
\end{verbatim}
depending on whether or not native-compiling is used. As a default, 
\texttt{lie.opt} is built; otherwise the user should edit the
\texttt{Makefile} correspondingly.

\newpage
\section{Test run description}
\label{s:test}
\subsection{Derivation of explicit lowering normalizations for arbitrary irreps}
\label{app:deriv_explicit_lows}
Bearing in mind that we only know consistent lowering normalizations
and scalar products of basis states for non-degenerate or adjoint
representations, see Section \ref{explicit_lowering}, we developed
means to derive also those for more complex irreps: As, obviously, any
irrep can be constructed in the (multiple) tensor product of
non-degenerate or adjoint representations, basis choices, scalar
products and their lowering normalizations can be analyzed therein. As
an example, we give the \texttt{O'Caml} toplevel code needed to create
an object associated to  the $\bf 650$ of $E_6$ which we know has many 
degeneracies. 

\begin{verbatim}
(* Create objects for the (anti-)fundamental E6 irreps, the 27 and 27bar *)
let e6_27    = new irrep E6 [1;0;0;0;0;0] empty_input;;
let e6_27bar = new irrep E6 [0;0;0;0;1;0] empty_input;;

(* Create object of the tensor product decomposition, 27x27bar *)
let cg = new clebsch_gordan_decomposition e6_27 e6_27bar;;
(* ... computes the first irrep in 27x27bar, the 650 *)
cg#decompose;;
(* ... computes all remaining irreps, yields:
Dimensions match.
Clebsch-Gordan decomposition successfully done!
E6: (1,0,0,0,0,0,)27 x (0,0,0,0,1,0,)27 = 
(1,0,0,0,1,0,)650
(0,0,0,0,0,1,)78
(0,0,0,0,0,0,)1
*)

(* Read out the 650 *)
let tmp = nth cg#get_irreps_in_product 0;;
(* Analyze lowerings and scalar products *)
let low_scp = tmp#prepare e6_27 e6_27bar;;
(* Create object associated to the 650 irrep using low_scp as input *)
let e6_650 = new irrep E6 [] low_scp;;
(* This object can now be used in further (multiple) tensor products *)
\end{verbatim}

\subsection{$\bf 248\otimes 248$ of $E_8$} 
By the very nature of the explicit decomposition algorithm, dimension
of tensor products are limited by memory and CPU time demands. While
the $\bf 650\otimes 650$ of $E_6$ exceeds 4 GB RAM, the software
successfully produces Clebsch-Gordan coefficients for $\bf 248\otimes
248$ of $E_8$. Input code in an \texttt{O'Caml} toplevel is as
follows: 

\begin{verbatim} 
(* Create object for the fundamental and adjoint 248 irrep *)
let e8_248 = new irrep E8 [0;0;0;0;0;0;1;0] empty_input;;

(* Create object of the tensor product decomposition *)
let cg = new clebsch_gordan_decomposition e8_248 e8_248;;
(* ... computes the first irrep in 248x248, the 27000... *)
cg#decompose;;
(* ... computes all remaining irreps *)
(* yields
Dimensions match.
Clebsch-Gordan decomposition successfully done!
E8: (0,0,0,0,0,0,1,0,)248 x (0,0,0,0,0,0,1,0,)248 = 
(0,0,0,0,0,0,2,0,)27000
(0,0,0,0,0,1,0,0,)30380
(1,0,0,0,0,0,0,0,)3875
(0,0,0,0,0,0,1,0,)248
(0,0,0,0,0,0,0,0,)1
*)

(* Get singlet contraction in tensor product *)
let singlet = nth cg#get_irreps_in_product 4;;
singlet#print e8_248 e8_248;;
(*
[[[("-1", ("(0,0,0,0,0,0,1,0,)1", "(0,0,0,0,0,0,-1,0,)1"));
  ("1", ("(0,0,0,0,0,1,-1,0,)1", "(0,0,0,0,0,-1,1,0,)1"));
  ("-1", ("(0,0,0,0,1,-1,0,0,)1", "(0,0,0,0,-1,1,0,0,)1"));
...]]]
*)
\end{verbatim} 

%%%%%

\subsection{Multiple tensor product: $\bf 4\otimes 4 \otimes 6 \otimes 
  15$ of $SU(4)$}
In what follows we present the \texttt{O'Caml} toplevel code needed to
compute the 4-fold tensor products of $SU(4)$ irreps,
\begin{equation}
  {\bf 4\otimes 4 \otimes 6 \otimes 15 = 1 + 1 + \ldots},
\end{equation}
where $\ldots$ denote further non-singlet irreps. Subsequently, we let
the ${\bf 15}$ acquire an $SU(3)\times U(1)$-invariant vev and
compute the resulting effective potential.

\begin{verbatim} 
(* Creating objects for irreps the tensor product is taken of *)
(* As all irreps are either non-degenerate or adjoint, consistent
lowering normalizations and scalar products can be generically computed *)
let su4_4    = new irrep (A 3) [1;0;0] empty_input;;
let su4_6    = new irrep (A 3) [0;1;0] empty_input;;
let su4_15   = new irrep (A 3) [1;0;1] empty_input;;

(* Creating objects for irreps that can be tensorized *)
let t4    = new irrep_in_tp su4_4;;
let t6    = new irrep_in_tp su4_6;;
let t15   = new irrep_in_tp su4_15;;

(* Define SU(3)xU(1)-singlet in 15 of SU(4) *)
let sing_su4_15 = ((new lcrs_vector [(unity_f, 7); (number (-2) 1 1,8);
                   (number 3 1 1, 9)])#scale (number 1 6 6))#simplify;;
(* Check singlet, by descending with simple roots 1 and 2 *)
(e_lower su4_15 1 sing_su4_15)#print;;
(e_lower su4_15 2 sing_su4_15)#print;;
(* In both cases, yields [], i.e. 0 *)

(* Further states of weight zero *)
let bs15 = [new lcrs_vector [(unity_f, 7+1)];
            (new lcrs_vector [(unity_f,7+2)])#scale (number 1 1 3)];;
(* Check linear independence of states in the zero weight space *)
linearly_dependent (sing_su4_15 :: bs15);;
(* Yields false, i.e. vectors are linearly independent *)
(* Using Gram-Schmidt procedure, choose basis in zero weight space
consisting of the SU(3)xU(1) singlet and two states orthogonal to this
singlet *)
let zero_ws_su4_15 = gram su4_15 [sing_su4_15] bs15;;
let basis_su4_15   = [sing_su4_15] @ zero_ws_su4_15;;

(* Compute transformation of the generic basis to the one in basis_su4_15  *)
let trafo = chbasis_list basis_su4_15 7;;

(* Compute all singlets in 4x4x6x15 *)
(* t4#otimes t4 1 yields the first irrep in the tensor product 4x4 *)
let tt1 = ((t4#otimes t4 1)#otimes t6 2)#otimes t15 7;;
(* ... evaluation of the tensor products *)
let tt2 = ((t4#otimes t4 2)#otimes t6 2)#otimes t15 7;;
(* ... evaluation of the tensor products *)

(* Check that tensor product yield singlets *)
tt1#get_res;;
tt2#get_res;;
(* Yields [(1, <obj>)] => OK *)
(* Check (anti)symmetry under exchange of first two irreps *)
tt1#is_sym 1 2;;
(* Yields 1, i.e. symmetric *)
tt2#is_sym 1 2;;
(* Yields -1, i.e. antisymmetric *)

(* Filter out zero weight components (labeled 7,8,9) in the 15 in the
4-fold tensor product *)
(* Change zero-weight basis in 15 *)
let tt1'   = tt1#filter 4 [7;8;9];;
let tt1''  = tt1'#chbasis 4 trafo;;
let tt2'   = tt2#filter 4 [7;8;9];;
let tt2''  = tt2'#chbasis 4 trafo;;
(* Filter out vev component denoted -1 *)
let tt1''' = tt1''#filter 4 [-1];;
let tt2''' = tt2''#filter 4 [-1];;

(* Scale to nice overall normalization *)
let tt1'''' = tt1'''#scale (number 1 2 3);;
let tt1'''' = tt1'''#scale (number 1 1 3);;

(* Print result *)
(snd (hd tt1''''#untree))#print;;
(* Result:
[("-1", "(((4,3),1),-1)"); ("-1", "(((3,4),1),-1)"); ("1", "(((4,2),2),-1)");
 ("1", "(((2,4),2),-1)"); ("-1", "(((4,1),4),-1)"); ("-1", "(((1,4),4),-1)")] *)
(snd (hd tt2''''#untree))#print;;
(* Result:
[("1", "(((1,3),5),-1)"); ("-1", "(((3,1),5),-1)"); ("-1", "(((1,2),6),-1)");
 ("1", "(((2,1),6),-1)"); ("1", "(((3,4),1),-1)"); ("-1", "(((4,3),1),-1)");
 ("-1", "(((2,4),2),-1)"); ("1", "(((4,2),2),-1)"); ("1", "(((1,4),4),-1)");
 ("-1", "(((4,1),4),-1)"); ("-1", "(((2,3),3),-1)"); ("1", "(((3,2),3),-1)")] *)

(* Weight labels *)
su4_4#print_list;;
(* yields:
[(1, "(1,0,0,)1"); (2, "(-1,1,0,)1"); (3, "(0,-1,1,)1"); (4, "(0,0,-1,)1")] *)

su4_6#print_list;;
(* yields:
[(1, "(0,1,0,)1"); (2, "(1,-1,1,)1"); (3, "(1,0,-1,)1");
 (4, "(-1,0,1,)1"); (5, "(-1,1,-1,)1"); (6, "(0,-1,0,)1")] *)
\end{verbatim}

\section{Acknowledgements}

This work has been partially supported by the German Research Society (DFG)
under grant number RE-2850/1-1. JRR likes to thank Karl-Hermann Neeb
for teaching him representation theory of Lie algebras in
mathematical rigorosity, and particularly Thorsten Ohl for
the continued fun and pleasure with \texttt{O'Caml} and the way how to
deal with lists and all that. 

\appendix

\section{Appendices}

\subsection{Dynkin Diagrams}
\label{sec:dynkin}

\begin{figure}
  \begin{center}
    \includegraphics{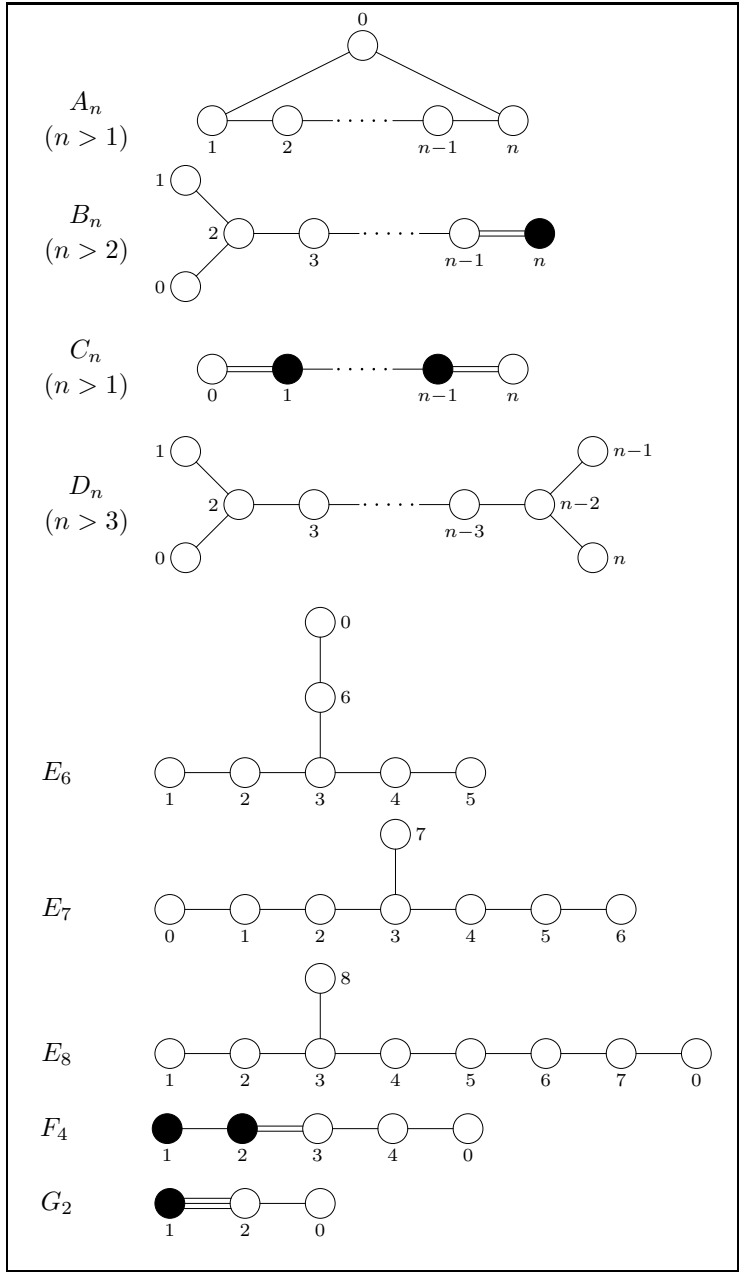}    
  \end{center}
\caption{\label{fig:dynkin} 
  (Extended) Dynkin diagrams with the labelling of simple roots and
  the lowest root according to our conventions. Black dots denote 
  shorter roots, as usual.}
\end{figure}

Our conventions for Dynkin diagrams are shown in Fig.~\ref{fig:dynkin}.

%%%%%

\subsection{Cartan Matrices}
\label{sec:cartan}

We follow the conventions by Georgi \cite{Georgi:1982jb} and Slansky
\cite{Slansky:1981yr} for the Cartan matrix, namely $A_{ji} = 2 \alpha^j \cdot
\alpha^i / (\alpha^i)^2$ which is the transpose of Cornwell
\cite{Cornwell}. For the Dynkin diagrams we follow the conventions of
\cite{Georgi:1982jb} concerning the normalization and numbering of the
weights, which has $n-1$ longer simple roots for $SO(2n+1)$ and $n-1$
shorter roots for $Sp(2n)$. For $F_4$ the first two roots on the left
of the Dynkin diagrams are the shorter ones, for $G_2$ the first root
on the left is the shorter, too. This is contrary to
\cite{Cornwell,Slansky:1981yr}, where the shorter roots are on the right hand
side. Therefore, the Cartan matrices for the infinite series and
$E_{6,7,8}$ given here agree with \cite{Slansky:1981yr} and are the transpose
for $F_4$, $G_2$, while the opposite is true for \cite{Cornwell}. 
Up to the definition of the Cartan matrix (which is important for the
descending from the highest weight) and the reflection of the Dynkin
diagrams for $F_4$ and $G_2$, we follow the conventions of Cornwell
\cite{Cornwell}. Our Cartan matrices are given by:

\begin{eqnarray}
 \label{eq:cartan1}
 A_{SU(n+1)} &=\; 
 \begin{scriptsize}
 \left(
 \begin{array}{rrrcrrr}
   2 & -1 & 0 & \hdots & 0 & 0 & 0 \\
   -1 & 2 & -1 & \hdots & 0 & 0 & 0 \\
   0 & -1 & 2 & \hdots & 0 & 0 & 0 \\
   \vdots & \vdots & \vdots & & 
   \vdots & \vdots & \vdots \\
   0 & 0 & 0 & \hdots & 2 & -1 & 0 \\
   0 & 0 & 0 & \hdots & -1 & 2 & -1 \\
   0 & 0 & 0 & \hdots & 0 & -1 & 2
 \end{array}
 \right),
 \end{scriptsize}
 \\
 A_{SO(2n+1)} &=\; 
 \begin{scriptsize}
 \left(
 \begin{array}{rrrcrrr}
   2 & -1 & 0 & \hdots & 0 & 0 & 0 \\
   -1 & 2 & -1 & \hdots & 0 & 0 & 0 \\
   0 & -1 & 2 & \hdots & 0 & 0 & 0 \\
   \vdots & \vdots & \vdots & & 
   \vdots & \vdots & \vdots \\
   0 & 0 & 0 & \hdots & 2 & -1 & 0 \\
   0 & 0 & 0 & \hdots & -1 & 2 & -2 \\
   0 & 0 & 0 & \hdots & 0 & -1 & 2
 \end{array}
 \right),  
 \end{scriptsize}
 \\
 A_{Sp(2n)} &=\; 
 \begin{scriptsize}
 \left(
 \begin{array}{rrrcrrr}
   2 & -1 & 0 & \hdots & 0 & 0 & 0 \\
   -1 & 2 & -1 & \hdots & 0 & 0 & 0 \\
   0 & -1 & 2 & \hdots & 0 & 0 & 0 \\
   \vdots & \vdots & \vdots & & 
   \vdots & \vdots & \vdots \\
   0 & 0 & 0 & \hdots & 2 & -1 & 0 \\
   0 & 0 & 0 & \hdots & -1 & 2 & -1 \\
   0 & 0 & 0 & \hdots & 0 & -2 & 2
 \end{array}
 \right),  
 \end{scriptsize}
 \\
 A_{SO(2n)} &=\;
 \begin{scriptsize}
 \left(
 \begin{array}{rrrcrrr}
   2 & -1 & 0 & \hdots & 0 & 0 & 0 \\
   -1 & 2 & -1 & \hdots & 0 & 0 & 0 \\
   0 & -1 & 2 & \hdots & 0 & 0 & 0 \\
   \vdots & \vdots & \vdots & & 
   \vdots & \vdots & \vdots \\
   0 & 0 & 0 & \hdots & 2 & -1 & -1 \\
   0 & 0 & 0 & \hdots & -1 & 2 & 0 \\
   0 & 0 & 0 & \hdots & -1 & 0 & 2
 \end{array}
 \right)
 \end{scriptsize}
\end{eqnarray}

\begin{multline}
 \label{eq:cartan2}
 A_{E_6} =
 \begin{scriptsize}
 \left(
 \begin{array}{rrrrrr}
          2 & -1 & 0 & 0 & 0 & 0  \\
          -1 & 2 & -1 & 0 & 0 & 0  \\
          0 & -1 & 2 & -1 & 0 & -1  \\
          0 & 0 & -1 & 2 & -1 & 0  \\
          0 & 0 & 0 & -1 & 2 & 0  \\
          0 & 0 & -1 & 0 & 0 & 2    
 \end{array}\right), \qquad
 A_{E_7} = 
 \left(
 \begin{array}{rrrrrrr}
          2 & -1 & 0 & 0 & 0 & 0 & 0  \\
          -1 & 2 & -1 & 0 & 0 & 0 & 0  \\
          0 & -1 & 2 & -1 & 0 & 0 & -1  \\
          0 & 0 & -1 & 2 & -1 & 0 & 0  \\
          0 & 0 & 0 & -1 & 2 & -1 & 0  \\
          0 & 0 & 0 & 0 & -1 & 2 & 0  \\
          0 & 0 & -1 & 0 & 0 & 0 & 2
 \end{array}\right), 
 \end{scriptsize}
 \\
 \begin{scriptsize}
 A_{E_8} = 
 \left(
 \begin{array}{rrrrrrrr}
          2 & -1 & 0 & 0 & 0 & 0 & 0 & 0  \\
          -1 & 2 & -1 & 0 & 0 & 0 & 0 & 0  \\
          0 & -1 & 2 & -1 & 0 & 0 & 0 & -1  \\
          0 & 0 & -1 & 2 & -1 & 0 & 0 & 0  \\
          0 & 0 & 0 & -1 & 2 & -1 & 0 & 0  \\
          0 & 0 & 0 & 0 & -1 & 2 & -1 & 0  \\
          0 & 0 & 0 & 0 & 0 & -1 & 2 & 0  \\
          0 & 0 & -1 & 0 & 0 & 0 & 0 & 2
 \end{array}\right), 
 \qquad
 A_{F_4} = 
 \left(
 \begin{array}{rrrr}
           2 & -1 & 0 & 0  \\
          -1 & 2 & -1 & 0  \\
          0 & -2 & 2 & -1  \\
          0 & 0 & -1 & 2    
 \end{array}\right),
 \end{scriptsize}
 \\
 \begin{scriptsize}
 A_{G_2} = 
 \left(
 \begin{array}{rr}
          2 & -1  \\
         -3 & 2    
 \end{array}
 \right)
 \end{scriptsize}
\end{multline}

%%%%%

%%%%%

\subsection{Positive roots}
\label{seq:positiveroots}

The positive roots are needed inside the Freudenthal formula for the
calculation of the degeneracy of the states in a representation. In
this section we denote as usual the simple roots by $\alpha^i$, $i =
1, \ldots, \text{rank}(\mathfrak{g})$. 

\begin{itemize}
 \item 
   $SU(n+1)$ has the $\tfrac12 n(n+1)$ positive roots:
   \begin{equation}
     \label{eq:pos_sun}
     \sum_{u = j}^k \alpha^u \qquad \text{with} \;
     j,k = 1, 2, \ldots, n \, ; \, j \leq k 
   \end{equation}

 \item
   \label{eq:pos_so2n+1}
   $SO(2n+1)$ has the $n^2$ positive roots:
    \begin{equation}
      \left\{
      \begin{array}{ll}
    	\sum_{u=j}^n \alpha^u & , \; j = 1,2,\ldots,n \\
    	\sum_{u=j}^{k-1} \alpha^u + 2 \sum_{u=k}^n \alpha^u &
	, \; j,k = 1,2, \ldots, n \, \; \, j < k \\
	\sum_{u=j}^{k-1} \alpha^u  &
	, \; j,k = 1,2, \ldots, n \, \; \, j < k
      \end{array}
      \right\}
    \end{equation}

  \item
   $Sp(2n)$ has the $n^2$ positive roots:
    \begin{equation}
   \label{eq:pos_sp2n}
      \left\{
      \begin{array}{ll}
    	\sum_{u=j}^{k-1} \alpha^u & , \; j,k = 1,2, \ldots, n \, ; \,
    	j < k \\ 
    	\sum_{u=j}^{k-1} \alpha^u + 2 \sum_{u=k}^{n-1} \alpha^u + 
	\alpha^n &
	, \; j,k = 1,2, \ldots, n-1 \, ; \, j < k \\	
	\sum_{u=j}^{n-1} \alpha^u  + \alpha^n &
	, \; j = 1,2, \ldots, n-1 
      \\
	2 \sum_{u=j}^{n-1} \alpha^u  + \alpha^n &
	, \; j = 1,2, \ldots, n-1 
      \\
      \alpha^n & 
      \end{array}
      \right\}
    \end{equation}

  \item
   $SO(2n)$ has the $n(n-1)$ positive roots:
    \begin{equation}
   \label{eq:pos_so2n}
      \left\{
      \begin{array}{ll}
    	\sum_{u=j}^{k-1} \alpha^u + 2 \sum_{u=k}^{n-2} \alpha^u + 
	\alpha^{n-1} + \alpha^n 
	& , \; j,k = 1,2, \ldots, n-2 \, ; \, j < k \, ; \, (n\geq 3)
	\\ 
    	\sum_{u=j}^{k-1} \alpha^u 
	& , \; j,k = 1,2, \ldots, n-2 \, ; \, j < k \, ; \, (n\geq 3)
	\\
	\sum_{u=j}^{n-2} \alpha^u  + \alpha^{n-1} + \alpha^n 
	& , \; j = 1,2, \ldots, n-2 \, ; \, (n\geq3)
	\\
	\sum_{u=j}^{n-2} \alpha^u  + \alpha^{n-1}
	& , \; j = 1,2, \ldots, n-2 \, ; \, (n\geq3)
	\\
	\sum_{u=j}^{n-2} \alpha^u  + \alpha^n
	& , \; j = 1,2, \ldots, n-2 \, ; \, (n\geq3)
	\\
	\sum_{u=j}^{n-2} \alpha^u 
	& , \; j = 1,2, \ldots, n-2 \, ; \, (n\geq3)
      \end{array}
      \right\}
    \end{equation}

  \item
    $E_6$ has 36 positive roots (coefficients of the simple roots):
    \begin{equation}
      \begin{scriptsize}
      \begin{array}{cccccc}
	   (100000), & (011100), & (111111)  &
	   (010000), & (001110), & (012101)  \\
	   (001000), & (001101), & (112101)  &
	   (000100), & (111100), & (012111)  \\
	   (000010), & (011110), & (112111)  &
	   (000001), & (001111), & (012211)  \\
	   (110000), & (111110), & (112211)  &
	   (011000), & (011001), & (122101)  \\
	   (001100), & (111001), & (122111)  &
	   (000110), & (011101), & (122211)  \\
	   (001001), & (111101), & (123211)  &
	   (111000), & (011111), & (123212)  
      \end{array}
      \end{scriptsize}
    \end{equation}

  \item
    $E_7$ has the 63 positive roots (coefficients of the simple
    roots):
    \begin{equation}
      \begin{scriptsize}
      \begin{array}{cccccc}
      (1000000), & (0001110), & (0122111)  & 
      (0100000), & (0011110), & (1122111)  \\ 
      (0010000), & (0111110), & (0122211)  & 
      (0001000), & (0011111), & (1122211)  \\
      (0000100), & (1111110), & (1221001)  & 
      (0000010), & (0110001), & (1221101)  \\ 
      (0000001), & (1110001), & (1221111)  & 
      (1100000), & (0111001), & (1222101)  \\ 
      (0110000), & (1111001), & (1222111)  & 
      (0011000), & (0111101), & (1222211)  \\ 
      (0010001), & (1111101), & (1232101)  &
      (1110000), & (0111111), & (1232111)  \\   
      (0001100), & (1111111), & (1232211)  &
      (0111000), & (0121001), & (1233211)  \\
      (0011001), & (1121001), & (1232102)  & 
      (1111000), & (0121101), & (1232112)  \\
      (0000110), & (1121101), & (1232212)  &
      (0011100), & (0121111), & (1233212)  \\
      (0111100), & (1121111), & (1243212)  &
      (0011101), & (0122101), & (1343212)  \\
      (1111100), & (1122101), & (2343212)  
      \end{array}
      \end{scriptsize}
    \end{equation}

  \item
    $E_8$ has the 120 positive roots (coefficients of the simple
    roots):     
    \begin{equation}
      \begin{scriptsize}
      \begin{array}{cccccc}
      (10000000), & (01111101), & (12332101)  &
      (01000000), & (11111101), & (12332111)  \\
      (00100000), & (01111111), & (12332211)  &
      (00010000), & (11111111), & (12333211)  \\
      (00001000), & (01210001), & (12321002)  &
      (00000100), & (11210001), & (12321102)  \\
      (00000010), & (01211001), & (12321112)  &
      (00000001), & (11211001), & (12322102)  \\
      (11000000), & (01211101), & (12322112)  &
      (01100000), & (11211101), & (12322212)  \\
      (00110000), & (01211111), & (12332102)  &
      (00100001), & (11211111), & (12332112)  \\
      (11100000), & (01221001), & (12332212)  &
      (00011000), & (11221001), & (12333212)  \\
      (01110000), & (01221101), & (12432102)  &
      (00110001), & (11221101), & (12432112)  \\
      (11110000), & (01221111), & (12432212)  &
      (00001100), & (11221111), & (12433212)  \\
      (00111000), & (01222101), & (12443212)  &
      (01111000), & (11222101), & (13432102)  \\
      (00111001), & (01222111), & (13432112)  &
      (11111000), & (11222111), & (13432212)  \\
      (00000110), & (01222211), & (13433212)  &
      (00011100), & (11222211), & (13443212)  \\
      (00111100), & (12210001), & (13543212)  &
      (01111100), & (12211001), & (13543213)  \\
      (00111101), & (12211101), & (23432102)  &
      (11111100), & (12211111), & (23432112)  \\
      (00001110), & (12221001), & (23432212)  &
      (00011110), & (12221101), & (23433212)  \\
      (00111110), & (12221111), & (23443212)  &
      (01111110), & (12222101), & (23543212)  \\
      (00111111), & (12222111), & (23543213)  &
      (11111110), & (12222211), & (24543212)  \\
      (01100001), & (12321001), & (24543213)  &
      (11100001), & (12321101), & (24643213)  \\
      (01110001), & (12321111), & (24653213)  &
      (11110001), & (12322101), & (24654213)  \\
      (01111001), & (12322111), & (24654313)  &
      (11111001), & (12322211), & (24654323)  	 
      \end{array}
      \end{scriptsize}      
    \end{equation}

  \item
    $F_4$ has the 24 positive roots (coefficients of the simple
    roots):     
    \begin{equation}
      \begin{scriptsize}
      \begin{array}{cccc}
	   (1000), & (0011), & (0121), & (1222)  \\
	   (0100), & (1110), & (1121), & (1231)  \\
	   (0010), & (0111), & (1220), & (1232)  \\
	   (0001), & (0120), & (0122), & (1242)  \\
	   (1100), & (1111), & (1122), & (1342)  \\
	   (0110), & (1120), & (1221), & (2342)  	 
      \end{array}
      \end{scriptsize}      
    \end{equation}

  \item
    $G_2$ has the 6 positive roots (coefficients of the simple
    roots):     
    \begin{equation}
      \begin{scriptsize}
      (10),  (01), (11),  (12), (13), (23)
      \end{scriptsize}
    \end{equation}
\end{itemize}

\newpage
%\section{References}
\bibliography{plain}

\end{document}